\documentclass[prd,preprintnumbers,twocolumn,amsmath,nofootinbib,amssymb]{revtex4}
\usepackage{graphicx,color,dcolumn,booktabs,bm}
\usepackage{longtable,lscape}
\usepackage{txfonts}
\usepackage{overpic}
\usepackage{amssymb}
\usepackage{epstopdf}
\usepackage{indentfirst}
\usepackage{feynmf}   
\usepackage{slashed}  
\usepackage{cases}
\usepackage{color}
\usepackage{float}
\usepackage{multirow}
\usepackage{ulem}
\usepackage{soul}
\usepackage{graphicx,color,dcolumn,booktabs,bm}
\usepackage{epsfig,dsfont,amssymb,amsmath,amsfonts,amsbsy,mathrsfs}

\graphicspath{{Figures/}} %

\usepackage{hyperref}
\hypersetup{colorlinks,citecolor=blue,anchorcolor=red,menucolor=red, linkcolor=red,filecolor=red,runcolor=red,urlcolor=blue,frenchlinks=red}

\makeatletter
\@addtoreset{equation}{section}
\makeatother

\allowdisplaybreaks

\begin{document}

\title{A new group of doubly charmed molecule with $T$-doublet charmed meson pair}
\author{Fu-Lai Wang$^{1,2}$}
\email{wangfl2016@lzu.edu.cn}
\author{Rui Chen$^{3}$}
\email{chenrui@hunnu.edu.cn}
\author{Xiang Liu$^{1,2,4}$\footnote{Corresponding author}}
\email{xiangliu@lzu.edu.cn}
\affiliation{$^1$School of Physical Science and Technology, Lanzhou University, Lanzhou 730000, China\\
$^2$Research Center for Hadron and CSR Physics, Lanzhou University and Institute of Modern Physics of CAS, Lanzhou 730000, China\\
$^3$Key Laboratory of Low-Dimensional Quantum Structures and Quantum Control of Ministry of Education, Department of Physics and Synergetic Innovation Center for Quantum Effectss and Applications, Hunan Normal University, Changsha 410081, China\\
$^4$Lanzhou Center for Theoretical Physics, Key Laboratory of Theoretical Physics of Gansu Province, and Frontiers Science Center for Rare Isotopes, Lanzhou University, Lanzhou 730000, China}

\begin{abstract}
Inspired by the observation of the doubly charmed tetraquark state $T_{cc}^+$ at the LHCb Collaboration, we systematically study the $S$-wave interactions between a pair of charmed mesons ($D_{1},\,D_{2}^{\ast}$) in the $T$-doublet, where we adopt the one-boson-exchange model and consider both the $S$-$D$ wave mixing and coupled channel effects. Our numerical results suggest that the prime doubly charmed molecular tetraquark candidates are the $S$-wave $D_1D_1$ state with $I(J^P)=0(1^+)$, the $S$-wave $D_1{D}_2^{\ast}$ states with $I(J^P)=0(1^+,\,2^+,\,3^+)$, and the $S$-wave ${D}_2^{\ast}{D}_2^{\ast}$ states with $I(J^P)=0(1^+,\,3^+)$. The $S$-wave $D_1D_1$ state with $I(J^P)=1(2^+)$, the $S$-wave $D_1{D}_2^{\ast}$ state with $I(J^P)=1(3^+)$, and the $S$-wave ${D}_2^{\ast}{D}_2^{\ast}$ state with $I(J^P)=1(4^+)$ may be the possible isovector doubly charmed molecular tetraquark candidates. We expect the experiments like the LHCb or Belle II collaborations to search for these predicted doubly charmed molecular tetraquarks near the $D_1D_1$, $D_1D_2^*$, and $D_2^*D_2^*$ thresholds in the future.
\end{abstract}

\maketitle

\section{Introduction}\label{sec1}

Very recent, the LHCb Collaboration reported the first observation of the doubly charmed tetraquark state $T_{cc}^+$ in the $D^0D^0\pi^+$ invariant mass spectrum \cite{LHCb:2021vvq}. Obviously, it is beyond the conventional hadron since the $T_{cc}^+$ has minimal quark component $cc\bar{u}\bar{d}$. In Ref. \cite{LHCb:2021auc}, the pole parameters $\hat{s}=m_{\text{pole}}-i\Gamma_{\text{pole}}/2$ for the $T_{cc}^+$ state related to the $D^0D^{*+}$ mass threshold are
\begin{eqnarray}
\delta m_{\text{pole}} &=& -360\pm 40^{+4}_{-0}~\text{keV}/c^{2},\nonumber\\
\Gamma_{\text{pole}} &=& 48\pm 2^{+0}_{-14}~\text{keV}.\nonumber
\end{eqnarray}
Additionally, its spin-parity quantum number is $J^P=1^+$. {After the observation of the $T_{cc}^+$ state, a series of theoretical studies tried to understand its inner structures and properties within the doubly charmed molecular tetraquark \cite{Li:2021zbw,Chen:2021vhg,Ren:2021dsi,Xin:2021wcr,Chen:2021tnn,Albaladejo:2021vln,Dong:2021bvy,Baru:2021ldu,Du:2021zzh,Kamiya:2022thy,Padmanath:2022cvl,Agaev:2022ast,Ke:2021rxd,Zhao:2021cvg,Deng:2021gnb,Santowsky:2021bhy,
Dai:2021vgf,Feijoo:2021ppq} and compact doubly charmed tetraquark pictures \cite{Guo:2021yws,Weng:2021hje,Kim:2022mpa,Agaev:2021vur}.}
{In fact, before the above LHCb experiment, several theoretical groups predicted the existence of the $DD^*$ doubly charmed molecular tetraquark state \cite{Manohar:1992nd,Ericson:1993wy,Tornqvist:1993ng,Janc:2004qn,Ding:2009vj,Molina:2010tx,Ding:2020dio,Li:2012ss,Xu:2017tsr,Liu:2019stu,Ohkoda:2012hv,Tang:2019nwv} and compact doubly charmed
tetraquark states \cite{Heller:1986bt,Carlson:1987hh,Silvestre-Brac:1993zem,Semay:1994ht,Moinester:1995fk,Pepin:1996id,Gelman:2002wf,Vijande:2003ki,Navarra:2007yw,Vijande:2007rf,Ebert:2007rn,Lee:2009rt,Yang:2009zzp,Luo:2017eub,
Karliner:2017qjm,Eichten:2017ffp,Wang:2017uld,Park:2018wjk,Junnarkar:2018twb,Deng:2018kly,Maiani:2019lpu,Yang:2019itm,Tan:2020ldi,Lu:2020rog,Braaten:2020nwp,Gao:2020ogo,Cheng:2020wxa,Noh:2021lqs,Faustov:2021hjs}, and suggested experimentalists to carry out the search for them.}

The importance of the observed $T_{cc}$ is equal to that of the $X(3872)$. In 2003, the $X(3872)$ discovered by the Belle Collaboration \cite{Choi:2003ue} inspired extensive discussion of assigning it as the $D\bar{D}^*$ charmoniumlike molecular state with $I(J^{PC})=0(1^{++})$  \cite{Wong:2003xk,Swanson:2003tb,Suzuki:2005ha,Liu:2008fh,Thomas:2008ja,Liu:2008tn,Lee:2009hy}. Obviously, it is a start point of exploring hidden-charm molecular tetraquark state since more hidden-charm molecular tetraquark explanations to the charmoniumlike $X/Y/Z$ states have been proposed \cite{Chen:2016qju,Liu:2013waa,Hosaka:2016pey,Liu:2019zoy,Brambilla:2019esw,Olsen:2017bmm,Guo:2017jvc}, which also stimulated the study of the hidden-charm molecular pentaquarks \cite{Wu:2010jy, Karliner:2015ina, Wang:2011rga, Yang:2011wz, Wu:2012md, Chen:2015loa} and the observation of the $P_c$ states \cite{Aaij:2015tga,Aaij:2019vzc} (see Table \ref{xyz} for a concise review).
\renewcommand\tabcolsep{0.00cm}
\renewcommand{\arraystretch}{1.50}
\begin{table}[!htbp]
\caption{A summary of experimental observations and the corresponding hidden-charm and doubly charmed molecular tetraquark explanations.}\label{xyz}
\begin{tabular}{l|l}
\toprule[1.0pt]\toprule[1.0pt]
\multicolumn{2}{c}{Hidden-charm molecular tetraquark systems}\\\midrule[1.0pt]
\multirow{3}{*}{$H\bar H$ \cite{Wang:2021aql,Wong:2003xk,Swanson:2003tb,Suzuki:2005ha,Liu:2008fh,Thomas:2008ja,Liu:2008tn,Lee:2009hy,Liu:2008mi,Liu:2009ei,Sun:2011uh,Sun:2012zzd,Zhao:2015mga,
Liu:2017mrh,Dong:2021juy,Liu:2016kqx,Dai:2018nmw,Yang:2017prf,Zhang:2006ix,Wang:2020elp,Dai:2020yfu}}&$X(3872)$ \cite{Choi:2003ue}, $Z_c(3900)$ \cite{Ablikim:2013mio}, $X(3915)$ \cite{Aubert:2007vj},\\
                                                                                                         & $Z(3930)$ \cite{Belle:2005rte}, $Y(3940)$ \cite{Abe:2004zs}, $Y(4008)$ \cite{Belle:2007dxy}, \\
                                                                                                         &$Z_c(4020)$ \cite{BESIII:2013ouc}, $Z_c(4025)$ \cite{BESIII:2013mhi}, $Z^+(4051)$ \cite{Mizuk:2008me}\\\midrule[0.5pt]
\multirow{2}{*}{$H\bar T$ \cite{Dong:2021juy,Close:2010wq,Li:2015exa,Close:2009ag,He:2017mbh,Li:2013bca,Zhu:2013sca}}&$Z^+(4248)$ \cite{Mizuk:2008me}, $Y(4260)$ \cite{Aubert:2005rm}, $Y(4360)$ \cite{BaBar:2006ait},\\
                                                                                                               &$Y(4390)$ \cite{BESIII:2016adj}, $Z^+(4430)$ \cite{Choi:2007wga}\\\midrule[0.5pt]
$H\bar S$ \cite{Shen:2010ky,Liu:2007bf,Liu:2008xz}&$Z^+(4248)$ \cite{Mizuk:2008me}, $Z^+(4430)$ \cite{Choi:2007wga}\\\midrule[0.5pt]
$T\bar T$ \cite{Dong:2021juy,Chen:2015add}&\\\midrule[0.5pt]
$S\bar S$ \cite{Hu:2010fg}&\\\midrule[1.0pt]
\multicolumn{2}{c}{Doubly charmed molecular tetraquark systems}\\\midrule[1.0pt]
$HH$ \cite{Li:2021zbw,Chen:2021vhg,Ren:2021dsi,Xin:2021wcr,Chen:2021tnn,Albaladejo:2021vln,Dong:2021bvy,Baru:2021ldu,Du:2021zzh,Kamiya:2022thy,Padmanath:2022cvl,Agaev:2022ast,Ke:2021rxd,Zhao:2021cvg,Deng:2021gnb,Santowsky:2021bhy,Manohar:1992nd,
Ericson:1993wy,Tornqvist:1993ng,Janc:2004qn,Ding:2009vj,Molina:2010tx,Ding:2020dio,Li:2012ss,Xu:2017tsr,Liu:2019stu,Ohkoda:2012hv,Tang:2019nwv}&$T_{cc}^+$ \cite{LHCb:2021vvq}\\\midrule[0.5pt]
$HT$ \cite{Wang:2021yld,Dong:2021bvy}&\\
\bottomrule[1.0pt]\bottomrule[1.0pt]
\end{tabular}
\end{table}

As we can see, there are several common features between the $X(3872)$ and $T_{cc}$ states. Their masses are very close to the threshold of the $DD^*$ pair, and the long-range interaction for the $D\bar{D}^*$ state with $I(J^{PC})=0(1^{++})$ is exactly same as the $DD^*$ state with $I(J^P)=0(1^+)$ \cite{Li:2021zbw}. Inspired by the $X(3872)$, a zoo of the hidden-charm molecular tetraquarks has been formed \cite{Wang:2021aql}. Along this line, the $T_{cc}^+$ observation must initiate a parallel zoo of the doubly charmed molecular tetraquarks. Searching for the doubly charmed molecular tetraquarks not only deepens our understanding to the interactions between charmed meson pair, but also may provide some hints to shed light on the interactions between charmed meson and anti-charmed meson.

In the heavy quark spin symmetry \cite{Wise:1992hn}, the charmed mesons with the same light quark spin-parity quantum number $j_l^P$ can be categorized into a doublet, such as $H[j_l^P=1/2^-] =(D,\,D^{\ast})$, $S[j_l^P=1/2^+] =(D_0^{\ast},\,D^{\prime}_1)$, and $T[j_l^P=3/2^+] =(D_1,\,D^{\ast}_2)$. {Here, we need to mention that the $D_1$ and $D^{\ast}_2$ states are abbreviation for the $D_1(2420)$ and $D^{\ast}_2(2460)$ states in the Particle Data Group (PDG), respectively.} In Refs. \cite{Manohar:1992nd,Ericson:1993wy,Tornqvist:1993ng,Janc:2004qn,Ding:2009vj,Molina:2010tx,Ding:2020dio,Li:2012ss,Xu:2017tsr,Liu:2019stu,Ohkoda:2012hv,Tang:2019nwv}, the authors predicted the possible doubly charmed molecular tetraquark candidates composed of a pair of charmed mesons in the $H$-doublet. Recently, we performed a systematic study of the $HT$-type meson-meson interactions, and predicted a series of possible $HT$-type doubly charmed molecular tetraquark candidates \cite{Wang:2021yld}.

{The newly reported doubly charmed tetraquark state $T_{cc}$ makes the study of the interactions between two charmed mesons an intriguing and important research issue \cite{LHCb:2021vvq,LHCb:2021auc}, we expect that there exists a zoo of the doubly charmed molecular tetraquarks in the hadron spectroscopy. In fact, the present research status of the doubly charmed molecular tetraquark states is similar to that of the hidden-charm molecular tetraquark states around 2003 \cite{Chen:2016qju}, the theorists should pay more attention to making the reliable prediction of various types of the doubly charmed molecular tetraquark states and give more abundant suggestions to searching for the doubly charmed molecular tetraquarks accessible at the forthcoming experiment, by which we hope that the crucial information can encourage experimental colleagues to focus on the doubly charmed molecular tetraquark states.}
For continuing the exploration of doubly charmed molecular tetraquark family, in this work, we focus on the $S$-wave interactions between a pair of charmed mesons in the $T$-doublet by adopting the one-boson-exchange (OBE) model. In realistic calculation, both the $S$-$D$ wave mixing effect and the coupled channel effect are considered. With these obtained OBE effective potentials, we can further solve the coupled channel Schr$\rm{\ddot{o}}$dinger equation to search for the bound state solutions. By this way, we study how charmed and anti-charmed mesons in the $T$ doublet form the $TT$-type doubly charmed molecular tetraquark states.
Finally, several possible $TT$-type doubly charmed molecular tetraquark candidates are predicted, which is the important step to construct the family of the doubly charmed molecular tetraquarks, where the provided crucial information is helpful to the experimental search for the possible doubly charmed molecular tetraquark candidates in the future expriment.

This paper is organized as follows. After the introduction in Sec. \ref{sec1}, we illustrate the detailed deduction of the $S$-wave interactions between a pair of charmed mesons in the $T$-doublet by adopting the OBE model in Sec. \ref{sec2}. With this preparation, we present the corresponding bound state properties of the $S$-wave $D_{1}D_{1}$, $D_1{D}_2^{\ast}$, and $D_{2}^{\ast}D_{2}^{\ast}$ systems, and further conclude whether there exist possible $TT$-type doubly charmed molecular tetraquark candidates in Sec. \ref{sec3}. This work ends with the summary in Sec. \ref{sec4}.

\section{Interactions of the $S$-wave $TT$ systems}\label{sec2}
In this section, we present the detailed derivation of the OBE effective potentials for these discussed $S$-wave $TT$-type doubly charmed tetraquark systems, and we include the contributions from the light scalar meson $(\sigma)$, pseudoscalar meson $(\pi/\eta)$, and vector meson $(\rho/\omega)$ exchange processes. {In this work, we treat the $D_1$ and $D_2^{*}$ states as the $P$-wave charmed mesons to study the doubly charmed molecular tetraquark candidates \cite{Chen:2016spr}. Here, we need to point out that the $D_1$ and $D_2^{*}$ states as the hadronic molecular states were also studied in the past decades \cite{Guo:2017jvc,Kolomeitsev:2003ac,Guo:2006rp,Gamermann:2007fi,Molina:2009eb}.}

\subsection{The flavor and spin-orbital wave functions}\label{subsec21}
Before deducing the effective potentials, we first construct the flavor and spin-orbital wave functions for these discussed $TT$-type doubly charmed tetraquark systems. The isospin quantum numbers for the $TT$-type doubly charmed tetraquark systems can be either $0$ or $1$, and we collect the relevant flavor wave functions $|I, I_3\rangle$ for the $TT$ systems in Table \ref{flavorwave} \cite{Li:2012ss,Chen:2021vhg,Liu:2019stu,Wang:2021yld}.
\renewcommand\tabcolsep{0.03cm}
\renewcommand{\arraystretch}{1.80}
\begin{table}[!htbp]
\caption{Flavor wave functions $|I, I_3\rangle$ for the $TT$ systems. Here, $I$ and $I_3$ stand for their isospin and the third component of these discussed $TT$ systems, respectively. For the $S$-wave $D_1{D}_2^{\ast}$ system, the factor $(-1)^{J_{D_1{D}_2^{\ast}}-3}$ comes from the exchange of two charmed mesons \cite{Liu:2013rxa,Chen:2015add,Zhu:2013sca,Wang:2021yld,Li:2015exa}.}\label{flavorwave}
\begin{tabular}{l|l|l}
\toprule[1.0pt]
\toprule[1.0pt]
\multicolumn{1}{l|}{$|I, I_3\rangle$}&$D_{1}D_{1}$ system &$D_{2}^{\ast}D_{2}^{\ast}$ system\\
\midrule[1.0pt]
$|1, 1\rangle$ &$|D_{1}^{+}D_{1}^{+}\rangle$   &$|D_{2}^{*+}D_{2}^{*+}\rangle$\\ $|1,0\rangle$&$\dfrac{1}{\sqrt{2}}\left(|D_{1}^{0}D_{1}^{+}\rangle+|D_{1}^{+}D_{1}^{0}\rangle\right)$\quad\quad\quad\quad
     &$\dfrac{1}{\sqrt{2}}\left(|D_{2}^{*0}D_{2}^{*+}\rangle+|D_{2}^{*+}D_{2}^{*0}\rangle\right)$\\
$|1, -1\rangle$&$|D_{1}^{0}D_{1}^{0}\rangle$
     &$|D_{2}^{*0}D_{2}^{*0}\rangle$\\
$|0,0\rangle$&$\dfrac{1}{\sqrt{2}}\left(|D_{1}^{0}D_{1}^{+}\rangle-|D_{1}^{+}D_{1}^{0}\rangle\right)$
     &$\dfrac{1}{\sqrt{2}}\left(|D_{2}^{*0}D_{2}^{*+}\rangle-|D_{2}^{*+}D_{2}^{*0}\rangle\right)$\\
\midrule[1.0pt]
\multicolumn{1}{l|}{$|I, I_3\rangle$}&\multicolumn{2}{l}{$D_1{D}_2^{\ast}$ system}\\
\midrule[1.0pt]
$|1, 1\rangle$ &\multicolumn{2}{l}{$\dfrac{1}{\sqrt{2}}\left(|D_{1}^{+}D_{2}^{*+}\rangle+(-1)^{J_{D_1{D}_2^{\ast}}-3}|D_{2}^{*+}D_{1}^{+}\rangle\right)$}\\
$|1,0\rangle$ &\multicolumn{2}{l}{$\dfrac{1}{2}\left[\left(|D_{1}^{0}D_{2}^{*+}\rangle+|D_{1}^{+}D_{2}^{*0}\rangle\right)
    +(-1)^{J_{D_1{D}_2^{\ast}}-3}\left(|D_{2}^{*+}D_{1}^{0}\rangle+|D_{2}^{*0}D_{1}^{+}\rangle\right)\right]$}\\
$|1, -1\rangle$
   &\multicolumn{2}{l}{$\dfrac{1}{\sqrt{2}}\left(|D_{1}^{0}D_{2}^{*0}\rangle+(-1)^{J_{D_1{D}_2^{\ast}}-3}|D_{2}^{*0}D_{1}^{0}\rangle\right)$}\\
$|0,0\rangle$    &\multicolumn{2}{l}{$\dfrac{1}{2}\left[\left(|D_{1}^{0}D_{2}^{*+}\rangle-|D_{1}^{+}D_{2}^{*0}\rangle\right)
    -(-1)^{J_{D_1{D}_2^{\ast}}-3}\left(|D_{2}^{*+}D_{1}^{0}\rangle-|D_{2}^{*0}D_{1}^{+}\rangle\right)\right]$}\\
\bottomrule[1.0pt]
\bottomrule[1.0pt]
\end{tabular}
\end{table}

These allowed quantum numbers for the $S$-wave $D_{1}D_{1}$ and $D_{2}^{\ast}D_{2}^{\ast}$ states include
\begin{eqnarray*}
&&D_1D_1[I(J^P)]: 1(0^+), 0(1^+), 1(2^+),\\
&&D_{2}^{\ast}D_{2}^{\ast}[I(J^P)]: 1(0^+), 0(1^+), 1(2^+), 0(3^+), 1(4^+).
\end{eqnarray*}
The spin-orbital wave functions $|{}^{2S+1}L_{J}\rangle$ for the $D_{1}D_{1}$, $D_1{D}_2^{\ast}$, and $D_{2}^{\ast}D_{2}^{\ast}$ systems can be expressed as
\begin{eqnarray*}
D_1D_1:&&|{}^{2S+1}L_{J}\rangle=\sum_{m,m^{\prime},m_S,m_L}C^{S,m_S}_{1m,1m^{\prime}}C^{J,M}_{Sm_S,Lm_L}\epsilon_{m}^\mu\epsilon_{m^{\prime}}^\nu|Y_{L,m_L}\rangle,\nonumber\\
D_1D_2^{\ast}:&&|{}^{2S+1}L_{J}\rangle=\sum_{m,m^{\prime},m_S,m_L}C^{S,m_S}_{1m,2m^{\prime}}C^{J,M}_{Sm_S,Lm_L}\epsilon_{m}^\lambda
\zeta_{m^{\prime}}^{\mu\nu}|Y_{L,m_L}\rangle,\nonumber\\
D_2^{\ast}D_2^{\ast}:&&|{}^{2S+1}L_{J}\rangle=\sum_{m,m^{\prime},m_S,m_L}C^{S,m_S}_{2m,2m^{\prime}}C^{J,M}_{Sm_S,Lm_L}
\zeta_{m}^{\mu\nu}\zeta_{m^{\prime}}^{\lambda\rho}|Y_{L,m_L}\rangle,
\end{eqnarray*}
respectively. Here, $C^{e,f}_{ab,cd}$ is the Clebsch-Gordan coefficient, and $|Y_{L,m_L}\rangle$ stands for the spherical harmonics function. $\epsilon^\mu_{m}\,(m=0,\,\pm1)$ and $\zeta^{\mu\nu}_{m}\,(m=0,\,\pm1,\,\pm2)$ denote the polarization vector and tensor for the axial-vector charmed meson $D_1$ and the tensor charmed meson $D_2^{*}$, respectively. In the static limit, their explicit expressions can be written as \cite{Cheng:2010yd}
\begin{eqnarray*}
\epsilon_{0}^{\mu}&=&\left(0,0,0,-1\right),~~~~~~\epsilon_{\pm}^{\mu}=\frac{1}{\sqrt{2}}\left(0,\,\pm1,\,i,\,0\right),\nonumber\\
\zeta^{\mu\nu}_{m}&=&\sum_{m_1,m_2}C^{2, m}_{1m_1,1m_2}\epsilon^{\mu}_{m_1}\epsilon^{\nu}_{m_2}.
\end{eqnarray*}

\subsection{The effective Lagrangians}\label{subsec22}
In the following, we adopt the effective Lagrangian approach to deduce the OBE effective potentials for the $D_{1}D_{1}$, $D_1{D}_2^{\ast}$, and $D_{2}^{\ast}D_{2}^{\ast}$ systems. Based on the heavy quark symmetry, the chiral symmetry, and the hidden gauge symmetry \cite{Casalbuoni:1992gi,Casalbuoni:1996pg,Yan:1992gz,Harada:2003jx,Bando:1987br}, the compact form of the relevant effective Lagrangians can be constructed as \cite{Ding:2008gr}
\begin{eqnarray}
\mathcal{L}_{TT\sigma}&=&g''_{\sigma}\langle T^{(Q)\mu}_a\sigma\overline{T}^{\,({Q})}_{a\mu}\rangle,\label{eq:lag1}\\
\mathcal{L}_{TT\mathbb{P}}&=&ik\langle T^{(Q)\mu}_{b}{\cal A}\!\!\!\slash_{ba}\gamma_5\overline{T}^{\,(Q)}_{a\mu}\rangle,\label{eq:lag2}\\
\mathcal{L}_{TT\mathbb{V}}&=&i\beta^{\prime\prime}\langle T^{(Q)\lambda}_bv^{\mu}({\cal V}_{\mu}-\rho_{\mu})_{ba}\overline{T}^{\,(Q)}_{a\lambda}\rangle\nonumber\\
&&+i\lambda^{\prime\prime}\langle T^{(Q)\lambda}_b\sigma^{\mu\nu}F_{\mu\nu}(\rho)_{ba}\overline{T}^{\,(Q)}_{a\lambda}\rangle,\label{eq:lag3}
\end{eqnarray}
where the four velocity is $v=(1, \bm{0})$ in the static approximation. Surperfield $T^{(Q)\mu}_a$ can be expressed as a linear combination of the axial-vector heavy flavor meson $P^{(Q)}_1$ with $I(J^P)=1/2(1^+)$ and the tensor heavy flavor meson $P_2^{*(Q)}$ with $I(J^P)=1/2(2^+)$, which reads as \cite{Ding:2008gr}
\begin{eqnarray*}
T^{(Q)\mu}_a&=&\frac{1+{v}\!\!\!\slash}{2}\left[P^{*(Q)\mu\nu}_{2a}\gamma_{\nu}-\sqrt{\frac{3}{2}}P^{(Q)}_{1a\nu}\gamma_5\left(g^{\mu\nu}
-\frac{\gamma^{\nu}(\gamma^{\mu}-v^{\mu})}{3}\right)\right],
\end{eqnarray*}
where $P_1^{(c)}=(D_1^0,\,D_1^+)$, $P_{2}^{*(c)}=(D_2^{*0},\,D_2^{*+})$, and the normalization relations for these discussed charmed mesons are $\langle 0|D_1^{\mu}|c\bar{q}(1^+)\rangle=\epsilon^{\mu}\sqrt{m_{D_1}}$ and $\langle 0|D_2^{*\mu\nu}|c\bar{q}(2^+)\rangle=\xi^{\mu\nu}\sqrt{m_{D_2^{\ast}}}$. Its conjugate field $\overline{T}_a^{(Q)\mu}$ is written as $\overline{T}_a^{(Q)\mu}=\gamma^0T_a^{(Q)\mu\dag}\gamma^0$.

In Eqs. (\ref{eq:lag1})-(\ref{eq:lag3}), we define the axial current $\mathcal{A}_\mu$, the vector current ${\cal V}_{\mu}$, the vector meson field $\rho_{\mu}$, and the vector meson field strength tensor $F_{\mu\nu}(\rho)$, i.e.,
\begin{eqnarray*}
\left.\begin{array}{ll}
{\mathcal A}_{\mu}=\dfrac{1}{2}(\xi^\dag\partial_\mu\xi-\xi \partial_\mu\xi^\dag),\quad\quad&{\mathcal V}_{\mu}=\dfrac{1}{2}(\xi^{\dagger}\partial_{\mu}\xi+\xi\partial_{\mu}\xi^{\dagger}),\\
\rho_{\mu}=\dfrac{i g_V}{\sqrt{2}}\mathbb{V}_{\mu},  &F_{\mu\nu}(\rho)=\partial_{\mu}\rho_{\nu}-\partial_{\nu}\rho_{\mu}+[\rho_{\mu},\rho_{\nu}],
\end{array}\right.
\end{eqnarray*}
with $\xi=\exp(i\mathbb{P}/f_\pi)$. Here, $\mathbb{P}$ and $\mathbb{V}_{\mu}$ stand for the pseudoscalar meson and vector meson matrices, respectively, which have the forms of
\begin{eqnarray*}
\left.\begin{array}{l}
{\mathbb{P}} = {\left(\begin{array}{ccc}
       \frac{\pi^0}{\sqrt{2}}+\frac{\eta}{\sqrt{6}} &\pi^+ &K^+\\
       \pi^-       &-\frac{\pi^0}{\sqrt{2}}+\frac{\eta}{\sqrt{6}} &K^0\\
       K^-         &\bar K^0   &-\sqrt{\frac{2}{3}} \eta     \end{array}\right)},\\
{\mathbb{V}}_{\mu} = {\left(\begin{array}{ccc}
       \frac{\rho^0}{\sqrt{2}}+\frac{\omega}{\sqrt{2}} &\rho^+ &K^{*+}\\
       \rho^-       &-\frac{\rho^0}{\sqrt{2}}+\frac{\omega}{\sqrt{2}} &K^{*0}\\
       K^{*-}         &\bar K^{*0}   & \phi     \end{array}\right)}_{\mu},
\end{array}\right.
\end{eqnarray*}
respectively. Once expanding the effective Lagrangians in Eqs. (\ref{eq:lag1})-(\ref{eq:lag3}), we can further obtain the concrete effective Lagrangians between the charmed mesons in the $T$-doublet and the light mesons, which can be explicitly expressed as
\begin{eqnarray}
\mathcal{L}_{D_1 D_1\sigma} &=& -2g_\sigma^{\prime\prime}D_{1a\mu}D^{\mu\dagger}_{1a} \sigma ,\\
\mathcal {L}_{D_1 D_1\mathbb{P}}&=&-\frac{5ik}{3f_\pi}~\epsilon^{\mu\nu\rho\tau}v_\tau D_{1b\nu}D^{\dagger}_{1a\mu} \partial_\rho\mathbb{P}_{ba},\\
\mathcal {L}_{D_1 D_1\mathbb{V}} &=& -\sqrt{2}\beta^{\prime \prime}g_{V}\left(v\cdot\mathbb{V}_{ba}\right) D_{1b\mu}D^{\mu\dagger}_{1a}\nonumber\\
&&+\frac{5\sqrt{2}i\lambda^{\prime\prime} g_{V}}{3}\left(D^{\mu}_{1b}D^{\nu\dagger}_{1a}-D^{\nu}_{1b}D^{\mu\dagger}_{1a}\right)\partial_\mu \mathbb{V}_{ba\nu},\\
\mathcal{L}_{D^{\ast}_2D^{\ast}_2\sigma} &=& 2g_\sigma^{\prime\prime}D^{*\mu\nu}_{2a}D^{*\dagger}_{2a\mu\nu}\sigma ,\\
\mathcal {L}_{D^{\ast}_2D^{\ast}_2\mathbb{P}}&=&\frac{2ik}{f_\pi}~\epsilon^{\mu\nu\rho\tau}v_\nu D^{*}_{2b\alpha\tau}D^{*\alpha\dagger}_{2a\rho}\partial_\mu\mathbb{P}_{ba},\\
\mathcal {L}_{D^{\ast}_2D^{\ast}_2\mathbb{V}} &=& \sqrt{2}\beta^{\prime \prime}g_{V}\left(v\cdot\mathbb{V}_{ba}\right) D_{2b}^{*\lambda\nu}  D^{*\dagger}_{2a{\lambda\nu}}+2\sqrt{2}i\lambda^{\prime\prime} g_{V}\nonumber\\
&&\times\left(D^{*\lambda\nu}_{2b} D^{*\mu\dagger}_{2a\lambda}-D^{*\mu}_{2b\lambda}D^{*\lambda\nu\dagger}_{2a}\right)\partial_\mu \mathbb{V}_{ba\nu},\\
\mathcal{L}_{D_1D^{\ast}_2\sigma} &=& \sqrt{\frac{2}{3}}ig_\sigma^{\prime\prime}\epsilon^{\mu \nu \rho \tau}v_{\rho}\left(D^{*}_{2a\mu\tau}D^{\dagger}_{1a\nu}-D_{1a\nu}D^{*\dagger}_{2a\mu\tau}\right) \sigma ,\\
\mathcal {L}_{D_1D^{\ast}_2\mathbb{P}}&=&-\sqrt{\frac{2}{3}}\frac{k}{f_\pi}\left(D^{*\mu\lambda}_{2b}D^{\dagger}_{1a\mu}
+D_{1b\mu}D^{*\mu\lambda\dagger}_{2a}\right)\partial_\lambda\mathbb{P}_{ba},\\
\mathcal {L}_{D_1D^{\ast}_2\mathbb{V}} &=& \frac{i\beta^{\prime \prime}g_{V}}{\sqrt{3}}\epsilon^{\lambda\alpha\rho\tau}v_{\rho}\left(v\cdot\mathbb{V}_{ba}\right)\left(D^{*}_{2b\lambda\tau}D^{\dagger}_{1a\alpha}-D_{1b\alpha}
D^{*\dagger}_{2a\lambda\tau}\right)\nonumber\\
&&+\frac{2\lambda^{\prime\prime} g_{V}}{\sqrt{3}}\left[3\epsilon^{\mu\lambda\nu\tau}v_\lambda\left(D^{*}_{2b\alpha\tau}D^{\alpha\dagger}_{1a}
+D^{\alpha}_{1b}D^{*\dagger}_{2a\alpha\tau}\right)\partial_\mu \mathbb{V}_{ba\nu}\right.\nonumber\\
&&+2\epsilon^{\lambda\alpha\rho\nu}v_\rho\left(D^{*\mu}_{2b\lambda}D^{\dagger}_{1a\alpha}+D_{1b\alpha}D^{*\mu\dagger}_{2a\lambda}\right)\nonumber\\
&&\left.\times\left(\partial_\mu \mathbb{V}_{ba\nu}-\partial_\nu \mathbb{V}_{ba\mu}\right)\right].
\end{eqnarray}

In this work, we estimate all the coupling constants in the quark model (see Refs. \cite{Wang:2019nwt,Liu:2011xc,Wang:2019aoc,Riska:2000gd} for more details). The values for the involved coupling constants are $g^{\prime\prime}_{\sigma}=0.76$,  $k=0.59$, $f_{\pi}=132~\rm{MeV}$, $\beta^{\prime\prime}=0.90$, $\lambda^{\prime\prime}=0.56~\rm{GeV}^{-1}$, and $g_V=5.83$ \cite{Wang:2019nwt}. The masses of these involved hadrons are $m_{\sigma}=600.00~\rm{MeV}$, $m_{\pi}=137.27~\rm{MeV}$, $m_{\eta}=547.86~\rm{MeV}$, $m_{\rho}=775.26~\rm{MeV}$, $m_{\omega}=782.66~\rm{MeV}$, $m_{D_{1}}=2422.00~\rm{MeV}$, and $m_{D_{2}^{*}}=2463.05~\rm{MeV}$ \cite{Zyla:2020zbs}. {As the important input parameters within the OBE model, the information about the coupling constants is crucial when studying the existence possibility of the hadronic molecular states. However, the coupling constants with the charmed meson in $T$-doublet cannot be fixed based on the experimental data at present, and we estimate these coupling constants by the quark model in this work \cite{Riska:2000gd}, which is often adopted to determine the coupling constants. Because of the uncertainties of the coupling constants, we should be cautious when studying the existence possibility of the hadronic molecular states. Here, we need to indicate our results of the existence possibility of the hadronic molecular states at the qualitative level, but the uncertainties of the coupling constants do not significantly affect our qualitative conclusions.}

\subsection{The effective potentials}\label{subsec23}

{Based on the above preparation, we can further deduce the effective potentials to judge whether these discussed doubly-charmed molecular tetraquark states exist or not. As indicated in Ref. \cite{Wang:2019nwt}, there exists the standard strategy to judge the possibility of the existence of the investigated hadronic molecular state within the OBE model, which is given in Fig. \ref{Procedure}.
\begin{figure}[!htbp]
\centering
\includegraphics[width=0.48\textwidth]{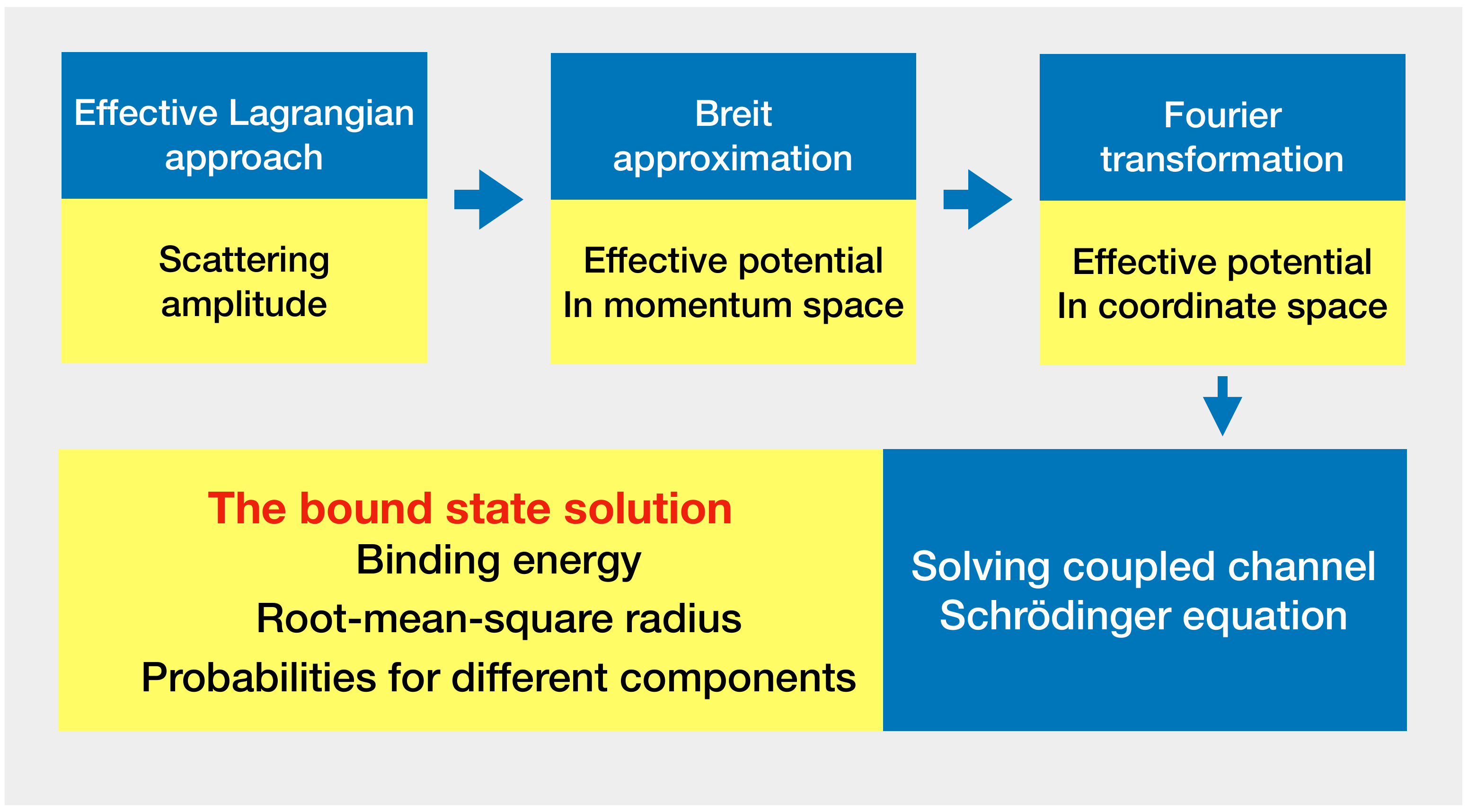}
\caption{(color online) The standard strategy to judge the possibility of the existence of the investigated hadronic molecular state within the OBE model.}\label{Procedure}
\end{figure}}

\begin{figure}[!htbp]
\centering
\includegraphics[width=0.23\textwidth]{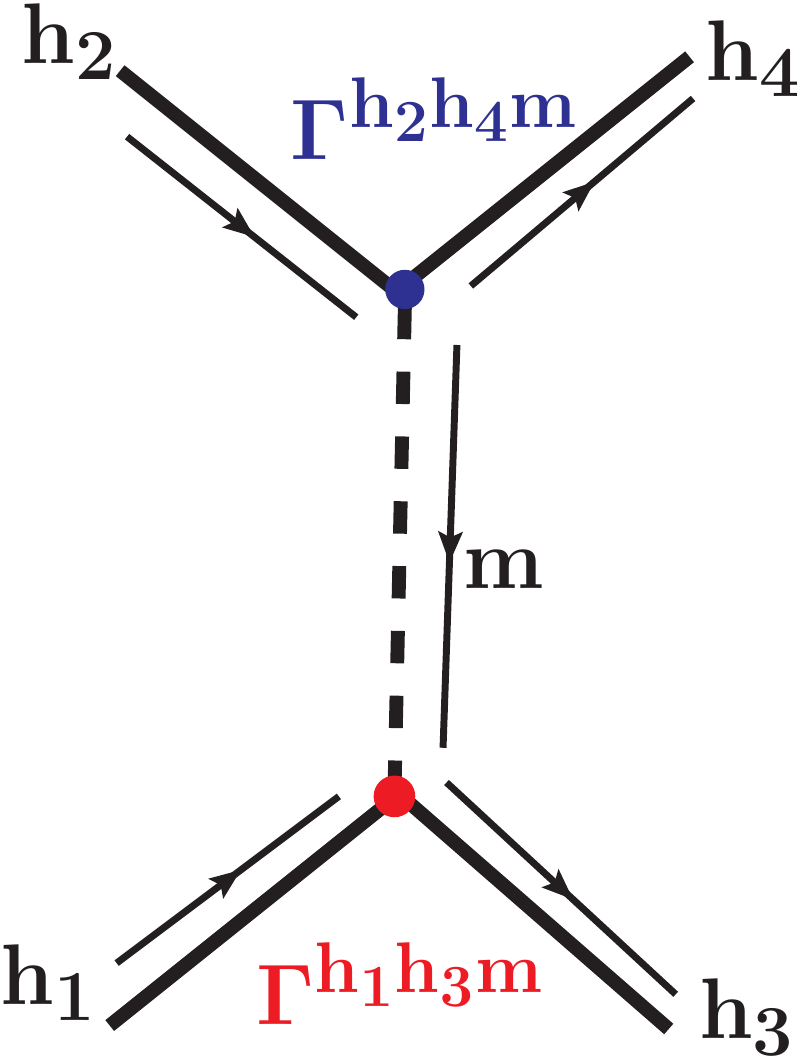}
\caption{The relevant Feynman diagram for the scattering process $h_1h_2\to h_3h_4$.}\label{fy}
\end{figure}

For extracting the effective potential in the coordinate space, which is shown in the upper part of the Fig. \ref{Procedure}, we will give a brief introduction in the following. First, we can write down the scattering amplitude $\mathcal{M}^{h_1h_2\to h_3h_4}(\bm{q})$ of the scattering process $h_1h_2\to h_3h_4$ by exchanging the light mesons under the effective Lagrangian approach. The relevant Feynman diagram is given in Fig.~\ref{fy}, and the scattering amplitude $\mathcal{M}^{h_1h_2\to h_3h_4}(\bm{q})$ can be written as
      \begin{eqnarray}\label{breit}
i\mathcal{M}^{h_1h_2\to h_3h_4}(\bm{q})=\sum_{m=\sigma,\,\mathbb{P},\,\mathbb{V}}i\Gamma^{h_1h_3m}_{(\mu)} P_{m}^{(\mu\nu)} i\Gamma^{h_2h_4m}_{(\nu)},
\end{eqnarray}
where the interaction vertices $\Gamma^{h_1h_3m}_{(\mu)}$ and $\Gamma^{h_2h_4m}_{(\nu)}$ can be extracted from the former effective Lagrangians.

And then, we adopt the Breit approximation \cite{Breit:1929zz,Breit:1930zza} and the nonrelativistic normalizations to obtain the relations between the effective potential in the momentum space $\mathcal{V}^{h_1h_2\to h_3h_4}_E(\bm{q})$ and the corresponding scattering amplitude $\mathcal{M}^{h_1h_2\to h_3h_4}(\bm{q})$, i.e.,
\begin{eqnarray}\label{breit}
\mathcal{V}_E^{h_1h_2\to h_3h_4}(\bm{q})=-\frac{\mathcal{M}^{h_1h_2\to h_3h_4}(\bm{q})}{\sqrt{\prod_i2m_i\prod_f2m_f}},
\end{eqnarray}
where $m_{i(f)}$ are the masses of the initial (final) states. Finally, we can perform the Fourier transformation to deduce the effective potential in the coordinate space $\mathcal{V}^{h_1h_2\to h_3h_4}_E(\bm{r})$, i.e.,
\begin{eqnarray}
\mathcal{V}^{h_1h_2\to h_3h_4}_E(\bm{r}) =\int\frac{d^3\bm{q}}{(2\pi)^3}e^{i\bm{q}\cdot\bm{r}}\mathcal{V}^{h_1h_2\to h_3h_4}_E(\bm{q})\mathcal{F}^2(q^2,m_E^2).\nonumber\\
\end{eqnarray}
In order to compensate the off-shell effect of the exchanged particles, we introduce the form factor $\mathcal{F}(q^2,m_E^2)$ at every interaction vertex, $\mathcal{F}(q^2,m_E^2) = (\Lambda^2-m_E^2)/(\Lambda^2-q^2)$. Here, $\Lambda$, $q$, and $m_E$ are the cutoff parameter, the four-momentum, and the mass of the exchanged light mesons, respectively. According to the experience of studying the deuteron, a reasonable cutoff value is taken around 1.00 GeV \cite{Tornqvist:1993ng,Tornqvist:1993vu}, in this cutoff region, we can reproduce the masses of the three $P_c$ states \cite{Wu:2010jy, Karliner:2015ina, Wang:2011rga, Yang:2011wz, Wu:2012md, Chen:2015loa} and the $T_{cc}$ state \cite{Li:2021zbw,Chen:2021vhg,Agaev:2021vur,Ren:2021dsi,Xin:2021wcr,Chen:2021tnn,Albaladejo:2021vln,Ding:2020dio,Li:2012ss,Xu:2017tsr,Liu:2019stu,Dong:2021bvy} in the hadronic molecular picture.

{After getting the effective potentials in the coordinate space for the discussed systems by the above three typical steps, we can search for the bound state solutions by numerically solving the coupled channel Schr\"{o}dinger equation,
\begin{eqnarray}
-\frac{1}{2\mu}\left(\nabla^2-\frac{\ell(\ell+1)}{r^2}\right)\psi(r)+V(r)\psi(r)=E\psi(r),
\end{eqnarray}
where $\nabla^2=\frac{1}{r^2}\frac{\partial}{\partial r}r^2\frac{\partial}{\partial r}$, and $\mu=\frac{m_1m_2}{m_1+m_2}$ is the reduced mass for the discussed systems. In our calculation, the obtained bound state solutions mainly include the binding energy $E$ and the radial wave function $\psi(r)$, with which we can further calculate the root-mean-square radius $r_{\rm RMS}$ and the probability of the individual channel $P_i$.}

The total OBE effective potentials for all the discussed $TT$-type doubly charmed tetraquarks can be expressed as
\begin{eqnarray}
\mathcal{V}_1&=&-g^{\prime\prime2}_{\sigma}\mathcal{O}_{1}Y_{\sigma}-\frac{25k^2}{108f^2_{\pi}}\left(\mathcal{O}_{2}\mathcal{Z}_r+\mathcal{O}_{3}\mathcal{T}_r\right)
\left(\frac{\mathcal{G}Y_{\pi}}{2}+\frac{Y_{\eta}}{6}\right)\nonumber\\
&&+\frac{1}{2}\beta^{\prime\prime2}g^2_V\mathcal{O}_{1}\left(\frac{\mathcal{G}Y_{\rho}}{2}+\frac{Y_{\omega}}{2}\right)\nonumber\\
&&-\frac{25}{54}\lambda^{\prime\prime2}g^2_V\left(2\mathcal{O}_{2}\mathcal{Z}_r-\mathcal{O}_{3}\mathcal{T}_r\right)\left(\frac{\mathcal{G}Y_{\rho}}{2}+\frac{Y_{\omega}}{2}\right)
\end{eqnarray}
for the ${D}_1{D}_1\to{D}_1{D}_1$ process,
\begin{eqnarray}
\mathcal{V}_2&=&-g^{\prime\prime2}_{\sigma}\frac{\mathcal{O}_{4}+\mathcal{O}_{4}^{\prime}}{2}Y_{\sigma}\nonumber\\
&&-\frac{5k^2}{18f^2_{\pi}}\left(\frac{\mathcal{O}_{5}+\mathcal{O}_{5}^{\prime}}{2}\mathcal{Z}_r+\frac{\mathcal{O}_{6}+\mathcal{O}_{6}^{\prime}}{2}\mathcal{T}_r\right)
\left(\frac{\mathcal{G}Y_{\pi}}{2}+\frac{Y_{\eta}}{6}\right)\nonumber\\
&&+\frac{1}{2}\beta^{\prime\prime2}g^2_V\frac{\mathcal{O}_{4}+\mathcal{O}_{4}^{\prime}}{2}\left(\frac{\mathcal{G}}{2}Y_{\rho}+\frac{1}{2}Y_{\omega}\right)\nonumber\\
&&-\frac{2}{3}\lambda^{\prime\prime2}g^2_V\left(2\frac{\mathcal{O}_{5}+\mathcal{O}_{5}^{\prime}}{2}\mathcal{Z}_r-\frac{\mathcal{O}_{6}
+\mathcal{O}_{6}^{\prime}}{2}\mathcal{T}_r\right)\left(\frac{\mathcal{G}Y_{\rho}}{2}+\frac{Y_{\omega}}{2}\right)\nonumber\\
\end{eqnarray}
for the ${D}_1{D}_2^{\ast}\to{D}_1{D}_2^{\ast}$ process,
\begin{eqnarray}
\mathcal{V}_3&=&\frac{k^2}{18f^2_{\pi}}\left(\frac{\mathcal{O}_{7}+\mathcal{O}_{7}^{\prime}}{2}\mathcal{Z}_r+\frac{\mathcal{O}_{8}
+\mathcal{O}_{8}^{\prime}}{2}\mathcal{T}_r\right)\left(\frac{\mathcal{G}Y_{\pi0}}{2}+\frac{Y_{\eta0}}{6}\right)\nonumber\\
&&+\frac{\lambda^{\prime\prime2}g^2_V}{9}\left(2\frac{\mathcal{O}_{7}+\mathcal{O}_{7}^{\prime}}{2}\mathcal{Z}_r-\frac{\mathcal{O}_{8}
+\mathcal{O}_{8}^{\prime}}{2}\mathcal{T}_r\right)\left(\frac{\mathcal{G}Y_{\rho0}}{2}+\frac{Y_{\omega0}}{2}\right)\nonumber\\
\end{eqnarray}
for the ${D}_1{D}_2^{\ast}\to{D}_2^{\ast}{D}_1$ process, and
\begin{eqnarray}
\mathcal{V}_4&=&-g^{\prime\prime2}_{\sigma}\mathcal{O}_{9}Y_{\sigma}-\frac{k^2}{3f^2_{\pi}}\left(\mathcal{O}_{10}\mathcal{Z}_r
+\mathcal{O}_{11}\mathcal{T}_r\right)\left(\frac{\mathcal{G}Y_{\pi}}{2}+\frac{Y_{\eta}}{6}\right)\nonumber\\
&&+\frac{1}{2}\beta^{\prime\prime2}g^2_V\mathcal{O}_{9}\left(\frac{\mathcal{G}Y_{\rho}}{2}+\frac{Y_{\omega}}{2}\right)\nonumber\\
&&-\frac{2}{3}\lambda^{\prime\prime2}g^2_V\left(2\mathcal{O}_{10}\mathcal{Z}_r-\mathcal{O}_{11}\mathcal{T}_r\right)\left(\frac{\mathcal{G}Y_{\rho}}{2}+\frac{Y_{\omega}}{2}\right)
\end{eqnarray}
for the ${D}_2^{\ast}{D}_2^{\ast}\to{D}_2^{\ast}{D}_2^{\ast}$ process. Here, the constant $\mathcal{G}$ is taken as $-3$ for the isoscalar system and 1 for the isovector system, respectively. For the convenience, we define the following expressions
\begin{eqnarray}
\mathcal{Z}_r&=&\frac{1}{r^2}\frac{\partial}{\partial r}r^2\frac{\partial}{\partial r},~~~~~~~~~~\mathcal{T}_r = r\frac{\partial}{\partial r}\frac{1}{r}\frac{\partial}{\partial r},\nonumber\\
Y_i&=&\dfrac{e^{-m_ir}-e^{-\Lambda_ir}}{4\pi r}-\dfrac{\Lambda_i^2-m_i^2}{8\pi\Lambda_i}e^{-\Lambda_ir}.
\end{eqnarray}
Here, $m_0=\sqrt{m^2-q_0^2}$ and $\Lambda_0=\sqrt{\Lambda^2-q_0^2}$ with $q_0 =m_{{D}_2^{\ast}}-m_{{D}_1}$.

In the above OBE effective potentials, we define several relevant operators $\mathcal{O}_k^{(\prime)}$, which include
\begin{eqnarray}\label{op}
\mathcal{O}_{1}&=&\left({\bm\epsilon^{\dagger}_3}\cdot{\bm\epsilon_1}\right)\left({\bm\epsilon^{\dagger}_4}\cdot{\bm\epsilon_2}\right), ~~~~~\mathcal{O}_{2}=\left({\bm\epsilon^{\dagger}_3}\times{\bm\epsilon_1}\right)\cdot\left({\bm\epsilon^{\dagger}_4}\times{\bm\epsilon_2}\right),\nonumber\\
\mathcal{O}_{3}&=&S({\bm\epsilon^{\dagger}_3}\times{\bm\epsilon_1},{\bm\epsilon^{\dagger}_4}\times{\bm\epsilon_2},\hat{\bm {r}}),\nonumber\\
\mathcal{O}_{4}&=&\mathcal{A}\left({\bm\epsilon^{\dagger}_3}\cdot{\bm\epsilon_1}\right)\left({\bm\epsilon^{\dagger}_{4a}}\cdot{\bm\epsilon_{2c}}\right)
\left({\bm\epsilon^{\dagger}_{4b}}\cdot {\bm\epsilon_{2d}}\right),\nonumber\\
\mathcal{O}_{4}^{\prime}&=&\mathcal{A}\left({\bm\epsilon^{\dagger}_4}\cdot{\bm\epsilon_2}\right)\left({\bm\epsilon^{\dagger}_{3a}}\cdot{\bm\epsilon_{1c}}\right)
\left({\bm\epsilon^{\dagger}_{3b}}\cdot {\bm\epsilon_{1d}}\right),\nonumber\\
\mathcal{O}_{5}&=&\mathcal{A}\left({\bm\epsilon^{\dagger}_{4a}}\cdot{\bm\epsilon_{2c}}\right)\left[\left({\bm\epsilon^{\dagger}_{3}}\times{\bm\epsilon_{1}}\right)
\cdot\left({\bm\epsilon^{\dagger}_{4b}}\times
{\bm\epsilon_{2d}}\right)\right],\nonumber\\
\mathcal{O}_{5}^{\prime}&=&\mathcal{A}\left({\bm\epsilon^{\dagger}_{3a}}\cdot{\bm\epsilon_{1c}}\right)\left[\left({\bm\epsilon^{\dagger}_{4}}\times{\bm\epsilon_{2}}\right)
\cdot\left({\bm\epsilon^{\dagger}_{3b}}\times {\bm\epsilon_{1d}}\right)\right],\nonumber\\
\mathcal{O}_{6}&=&\mathcal{A}\left({\bm\epsilon^{\dagger}_{4a}}\cdot{\bm\epsilon_{2c}}\right)S({\bm\epsilon^{\dagger}_{3}}\times{\bm\epsilon_{1}},{\bm\epsilon^{\dagger}_{4b}}\times {\bm\epsilon_{2d}},\hat{\bm {r}}),\nonumber\\
\mathcal{O}_{6}^{\prime}&=&\mathcal{A}\left({\bm\epsilon^{\dagger}_{3a}}\cdot{\bm\epsilon_{1c}}\right)
S({\bm\epsilon^{\dagger}_{4}}\times{\bm\epsilon_{2}},{\bm\epsilon^{\dagger}_{3b}}\times {\bm\epsilon_{1d}},\hat{\bm {r}}),\nonumber\\
\mathcal{O}_{7}&=&\mathcal{A}\left({\bm\epsilon^{\dagger}_{3a}}\cdot{\bm\epsilon_1}\right)\left({\bm\epsilon^{\dagger}_{4}}\cdot{\bm\epsilon_{2c}}\right)
\left({\bm\epsilon^{\dagger}_{3b}}\cdot {\bm\epsilon_{2d}}\right),\nonumber\\
\mathcal{O}_{7}^{\prime}&=&\mathcal{A}\left({\bm\epsilon^{\dagger}_{4a}}\cdot{\bm\epsilon_2}\right)\left({\bm\epsilon^{\dagger}_{3}}\cdot{\bm\epsilon_{1c}}\right)
\left({\bm\epsilon^{\dagger}_{4b}}\cdot {\bm\epsilon_{1d}}\right),\nonumber\\
\mathcal{O}_{8}&=&\mathcal{A}\left({\bm\epsilon^{\dagger}_{3a}}\cdot{\bm\epsilon_1}\right)\left({\bm\epsilon^{\dagger}_{4}}\cdot{\bm\epsilon_{2c}}\right)
S({\bm\epsilon^{\dagger}_{3b}},{\bm\epsilon_{2d}},\hat{\bm {r}}),\nonumber\\
\mathcal{O}_{8}^{\prime}&=&\mathcal{A}\left({\bm\epsilon^{\dagger}_{4a}}\cdot{\bm\epsilon_2}\right)\left({\bm\epsilon^{\dagger}_{3}}\cdot{\bm\epsilon_{1c}}\right)
S({\bm\epsilon^{\dagger}_{4b}},{\bm\epsilon_{1d}},\hat{\bm {r}}),\nonumber\\
\mathcal{O}_{9}&=&\mathcal{B}\left({\bm\epsilon^{\dagger}_{3a}}\cdot{\bm\epsilon_{1c}}\right)\left({\bm\epsilon^{\dagger}_{3b}}\cdot{\bm\epsilon_{1d}}\right)
\left({\bm\epsilon^{\dagger}_{4e}}\cdot{\bm\epsilon_{2g}}\right)\left({\bm\epsilon^{\dagger}_{4f}}\cdot{\bm\epsilon_{2h}}\right),\nonumber\\
\mathcal{O}_{10}&=&\mathcal{B}\left({\bm\epsilon^{\dagger}_{3a}}\cdot{\bm\epsilon_{1c}}\right)\left({\bm\epsilon^{\dagger}_{4e}}\cdot{\bm\epsilon_{2g}}\right)
\left[\left({\bm\epsilon^{\dagger}_{3b}}\times{\bm\epsilon_{1d}}\right)\cdot\left({\bm\epsilon^{\dagger}_{4f}}\times{\bm\epsilon_{2h}}\right)\right],\nonumber\\
\mathcal{O}_{11}&=&\mathcal{B}\left({\bm\epsilon^{\dagger}_{3a}}\cdot{\bm\epsilon_{1c}}\right)\left({\bm\epsilon^{\dagger}_{4e}}\cdot{\bm\epsilon_{2g}}\right)
S({\bm\epsilon^{\dagger}_{3b}}\times{\bm\epsilon_{1d}},{\bm\epsilon^{\dagger}_{4f}}\times{\bm\epsilon_{2h}},\hat{\bm {r}})\nonumber\\
\end{eqnarray}
with $S({\bm x},{\bm y},\hat{\bm {r}})= 3\left(\hat{\bm r} \cdot {\bm x}\right)\left(\hat{\bm r} \cdot {\bm y}\right)-{\bm x} \cdot {\bm y}$, $\mathcal{A}=\sum_{a,b}^{c,d}C^{2,a+b}_{1a,1b}C^{2,c+d}_{1c,1d}$, and $\mathcal{B}=\sum_{a,b,c,d}^{e,f,g,h}C^{2,a+b}_{1a,1b}C^{2,c+d}_{1c,1d}C^{2,e+f}_{1e,1f}C^{2,g+h}_{1g,1h}$. The operator matrix elements $\langle f|\mathcal{O}_k^{(\prime)}|i\rangle$ are collected in Appendix \ref{app01}.

\section{Numerical results and discussions}\label{sec3}

In this section, we solve the coupled channel Schr$\rm{\ddot{o}}$dinger equations to search for the bound state solutions for the $S$-wave $D_1D_1$, $D_1D_2^*$, and $D_2^*D_2^*$ systems. Our numerical calculations are presented in three cases. First, we perform a single channel analysis with the OBE effective potentials. After that, we further introduce the $S$-$D$ wave mixing effect and the coupled channel effect, and give the numerical analysis again. With these steps, we can reveal the roles of the $S$-$D$ wave mixing effect and the coupled channel effect in the formation of the $TT$-type doubly charmed molecular tetraquark states.

When we consider the $S$-$D$ wave mixing effect and the coupled channel effect, the relevant spin-orbit wave functions $|^{2S+1}L_J\rangle$ for the $D_1D_1$, $D_1D_2^*$, and $D_2^*D_2^*$ systems are
\begin{eqnarray}
(I,J^P)=(1,0^+):&& D_{1}D_{1}|{}^1\mathbb{S}_0\rangle/|{}^5\mathbb{D}_0\rangle,~~ D_{2}^{\ast}D_{2}^{\ast}|{}^1\mathbb{S}_0\rangle/|{}^5\mathbb{D}_0\rangle;\nonumber\\
(I,J^P)=(0,1^+):&& D_{1}D_{1}|{}^3\mathbb{S}_1\rangle/|{}^3\mathbb{D}_1\rangle,~~ D_{1}D_{2}^{\ast}|{}^3\mathbb{S}_1\rangle/|{}^{3,5,7}\mathbb{D}_1\rangle,\nonumber\\
                && D_{2}^{\ast}D_{2}^{\ast}|{}^3\mathbb{S}_1\rangle/|{}^{3,7}\mathbb{D}_1\rangle;\nonumber\\
(I,J^P)=(1,1^+):&&  D_{1}D_{2}^{\ast}|{}^3\mathbb{S}_1\rangle/|{}^{3,5,7}\mathbb{D}_1\rangle;\nonumber\\
(I,J^P)=(0,2^+):&&  D_{1}D_{2}^{\ast}|^5\mathbb{S}_2\rangle/|{}^{3,5,7}\mathbb{D}_2\rangle;\nonumber\\
(I,J^P)=(1,2^+):&& D_{1}D_{1}|{}^5\mathbb{S}_2\rangle/|{}^{1,5}\mathbb{D}_2\rangle,~~ D_{1}D_{2}^{\ast}|^5\mathbb{S}_2\rangle/|{}^{3,5,7}\mathbb{D}_2\rangle,\nonumber\\
                && D_{2}^{\ast}D_{2}^{\ast}|{}^5\mathbb{S}_2\rangle/|^{1,5,9}\mathbb{D}_2\rangle;\nonumber\\
(I,J^P)=(0,3^+):&& D_{1}D_{2}^{\ast}|{}^7\mathbb{S}_3\rangle/|{}^{3,5,7}\mathbb{D}_{3}\rangle,~~ D_{2}^{\ast}D_{2}^{\ast}|{}^7\mathbb{S}_3\rangle/|{}^{3,7}\mathbb{D}_3\rangle;\nonumber\\
(I,J^P)=(1,3^+):&& D_{1}D_{2}^{\ast}|{}^7\mathbb{S}_3\rangle/|{}^{3,5,7}\mathbb{D}_{3}\rangle;\nonumber\\
(I,J^P)=(1,4^+):&& D_{2}^{\ast}D_{2}^{\ast}|{}^9\mathbb{S}_4\rangle/|{}^{5,9}\mathbb{D}_4\rangle.\nonumber
\end{eqnarray}

{Within the OBE model, the bound state properties are depend on the cutoff values, but the cutoff parameter cannot be determined exactly without relevant experiment data. According to the experience of studying the deuteron, we believe the cutoff parameter around 1.0 GeV is the reasonable input \cite{Tornqvist:1993ng,Tornqvist:1993vu}. Thus, we attempt to find the loosely bound state solutions of the $S$-wave $TT$ systems by varying the cutoff parameter in the range around 1.0 GeV in this work. In addition, when judging whether the bound state is an ideal hadronic molecular candidate, we expect that the typical root mean square radius should be larger than the size of all the included component hadrons, and the reasonable binding energy is around several MeV to several tens MeV only, which is due to the hadronic molecular state is a loosely bound state \cite{Chen:2016qju}. Therefore, we consider the root mean square radius is greater than 0.7 fm and the binding energy is less than $20.0$ MeV when presenting the loosely bound state solutions of the $S$-wave $TT$ systems in the present work. In short, a loosely bound state with the cutoff parameter closed to 1.0 GeV can be suggested as a good hadronic molecular candidate in realistic calculation.}

\subsection{The $S$-wave $D_{1}D_{1}$ system}\label{subsec31}

When we take the cutoff value $\Lambda$ in the range of $0.8<\Lambda<2.5$ GeV, we cannot find the bound state solutions for the $S$-wave $D_{1}D_{1}$ state with $I(J^P)=1(0^+)$, even if we include the $S$-$D$ wave mixing effect and the coupled channel effect.

\begin{figure}[!htbp]
\centering
\includegraphics[width=0.47\textwidth]{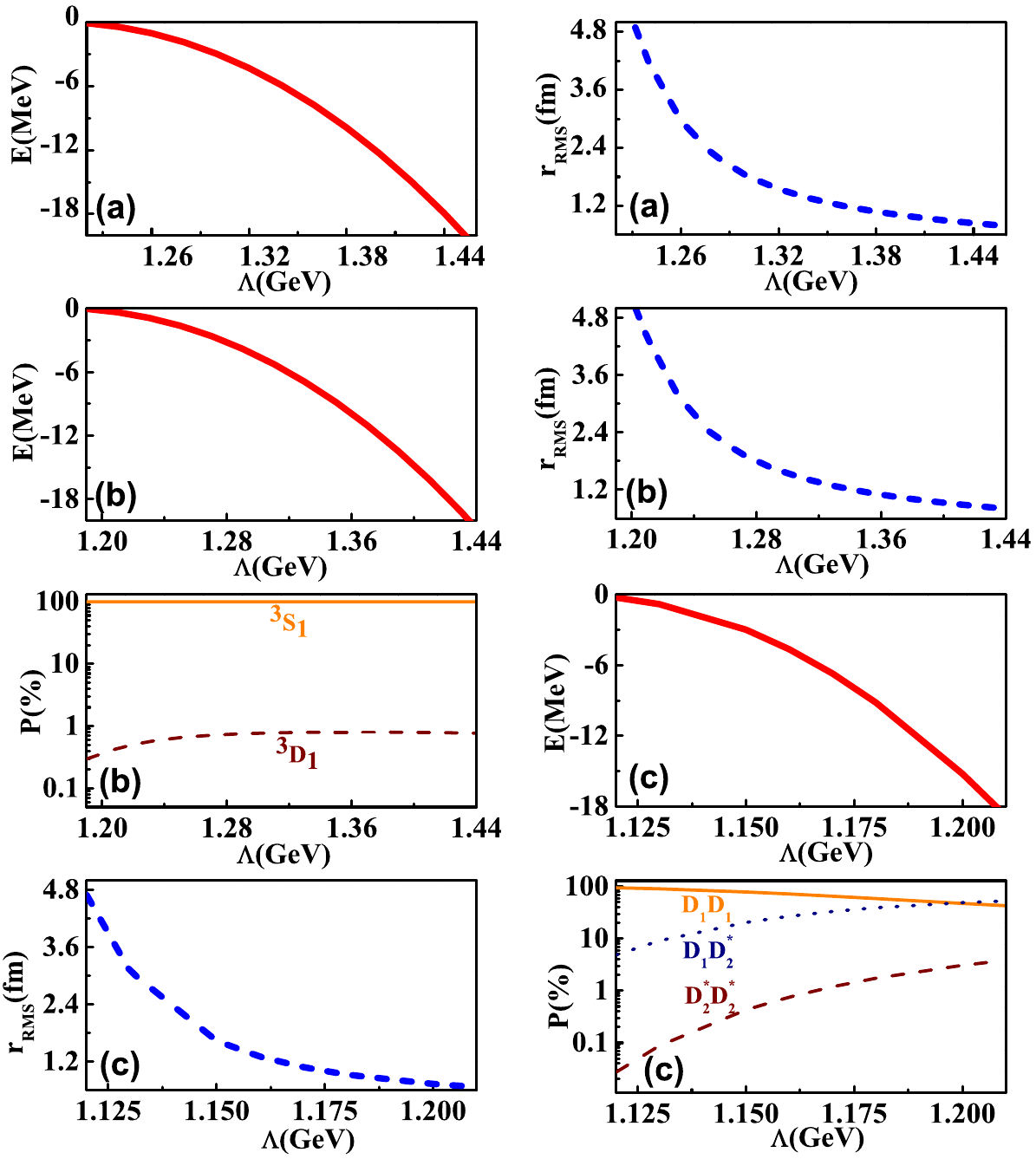}\\
\caption{(color online) The cutoff parameter $\Lambda$ dependence of the bound state solutions for the $S$-wave $D_{1}D_{1}$ state with $I(J^P)=0(1^+)$ when performing (a) the single channel analysis, (b) the $S$-$D$ wave mixing analysis, and (c) the coupled channel analysis.}\label{D1D11}
\end{figure}

In Fig.~\ref{D1D11}, we present the cutoff parameter $\Lambda$ dependence of the binding energy $E$, the root-mean-square (RMS) radius $r_{RMS}$, and the probabilities for different components $P_i$ for the $S$-wave $D_{1}D_{1}$ state with $I(J^P)=0(1^+)$ after considering the single channel, the $S$-$D$ wave mixing effect, and the coupled channel effect. As we can see, for the single $|{}^3\mathbb{S}_1\rangle$ channel analysis, the binding energy appears when taking the cutoff value $\Lambda \sim 1.23$ GeV. With increasing the cutoff value, its binding becomes  deeper. When the cutoff value $\Lambda$ is taken as 1.44 GeV, the binding energy increases to be 18.00 MeV, and the RMS radius decreases to be 0.80 fm. Thus, we suggest that the $S$-wave $D_{1}D_{1}$ state with $I(J^P)=0(1^+)$ can be a good doubly charmed molecular tetraquark candidate when we only consider the contribution of the $|{}^3\mathbb{S}_1\rangle$ channel. After considering the $S$-$D$ wave mixing effect, the bound state solutions are very similar to those in the single channel analysis, but the cutoff value $\Lambda$ becomes smaller if obtaining the same binding energy. Thus, the $S$-$D$ wave mixing effect is helpful to form the $D_{1}D_{1}$ bound state with $I(J^P)=0(1^+)$. Here, we find that the probability for the $D$-wave component for this bound state is very tiny, which is less than 1\%. When we perform the coupled channel analysis from the $D_{1}D_{1}|{}^3\mathbb{S}_1\rangle$, $D_1{D}_2^{\ast}|{}^3\mathbb{S}_1\rangle$, and $D_{2}^{\ast}D_{2}^{\ast}|{}^3\mathbb{S}_1\rangle$ interactions, we can obtain the loosely bound state solutions with the cutoff value $\Lambda$ around 1.12 GeV, which is a little smaller than those in the single channel and the $S$-$D$ wave mixing analysis. Obviously, the coupled channel effect can help to generate this bound state. When the binding energy is less than a dozen of MeV, the dominant channel is the $D_{1}D_{1}|{}^3\mathbb{S}_1\rangle$ component, followed by the $D_1{D}_2^{\ast}|{}^3\mathbb{S}_1\rangle$ and $D_{2}^{\ast}D_{2}^{\ast}|{}^3\mathbb{S}_1\rangle$ channels. As the increasing of the cutoff value, the $D_1{D}_2^{\ast}|{}^3\mathbb{S}_1\rangle$ and $D_{2}^{\ast}D_{2}^{\ast}|{}^3\mathbb{S}_1\rangle$ channels become much more important. To summary, the $S$-wave $D_{1}D_{1}$ state with $I(J^P)=0(1^+)$ can be a good doubly charmed molecular tetraquark candidate, and the $S$-$D$ wave mixing effect and the coupled channel effect play a minor role in generating the $S$-wave $D_{1}D_{1}$ bound state with $I(J^P)=0(1^+)$.

\begin{figure}[!htbp]
\centering
\includegraphics[width=0.47\textwidth]{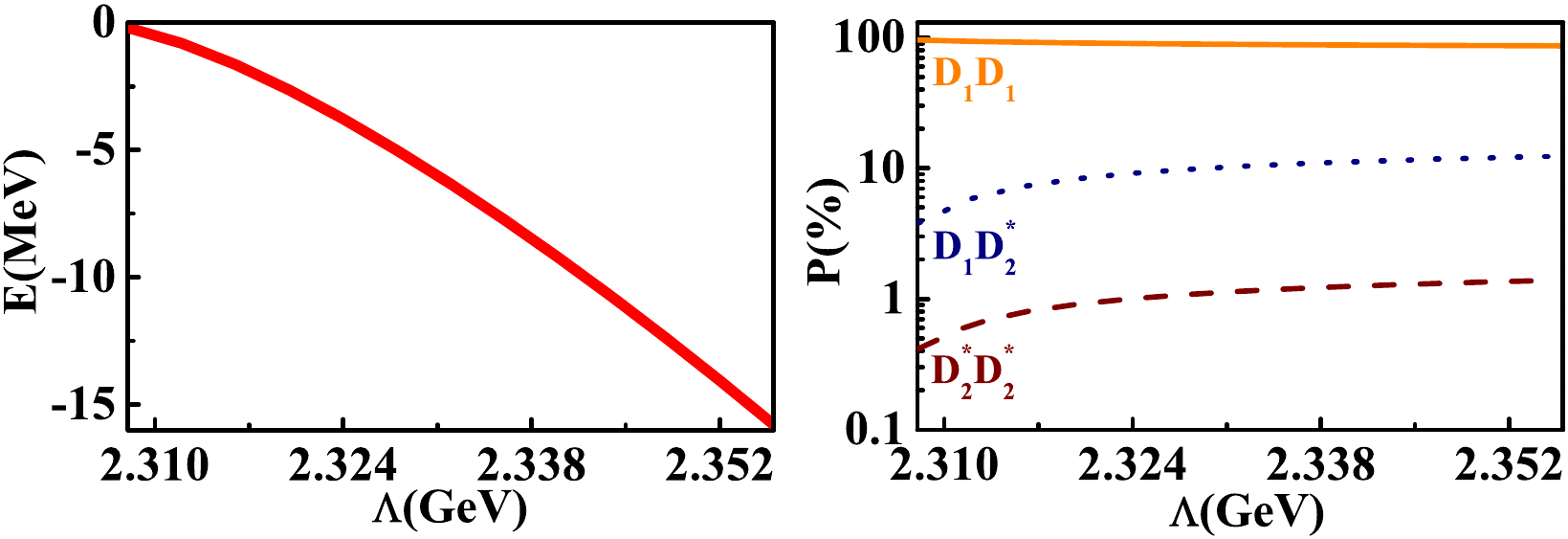}\\
\caption{The cutoff parameter $\Lambda$ dependence of the bound state solutions for the $S$-wave $D_{1}D_{1}$ state with $I(J^P)=1(2^+)$ by performing the coupled channel analysis.}\label{D1D12}
\end{figure}

For the $S$-wave $D_{1}D_{1}$ state with $I(J^P)=1(2^+)$, there does not exist the bound state solutions with the cutoff parameter around 0.80 to 2.50 GeV after performing both of the single channel analysis and the $S$-$D$ wave mixing analysis. However, when we take into account the coupled channel effect by including the $D_{1}D_{1}|{}^5\mathbb{S}_2\rangle$, $D_1{D}_2^{\ast}|{}^5\mathbb{S}_2\rangle$, and $D_{2}^{\ast}D_{2}^{\ast}|{}^5\mathbb{S}_2\rangle$ channels, we can find weakly binding with the cutoff value larger than 2.30 GeV, which is much larger than the cutoff value in the $S$-wave $D_{1}D_{1}$ state with $I(J^P)=0(1^+)$. Thus, the $S$-wave $D_{1}D_{1}$ state with $I(J^P)=1(2^+)$ may be the possible doubly charmed molecular tetraquark candidate. As shown in Fig.~\ref{D1D12}, we present the bound state solutions for the $S$-wave $D_{1}D_{1}$ state with $I(J^P)=1(2^+)$ after considering the coupled channel effects, where the $D_{1}D_{1}|{}^5\mathbb{S}_2\rangle$ is the dominant channel with the probability over 80\%.

{Because the hadrons are not point-like particles, we usually introduce the form factor at each interaction vertexes to describe the off-shell effect of the exchanged light mesons and reflect the inner structure effect of the discussed hadrons. As a general rule, the monopole form factor $\mathcal{F}(q^2,m_E^2) = (\Lambda^2-m_E^2)/(\Lambda^2-q^2)$ is often adopted to discuss the hadron-hadron interactions within the OBE model, which may reduce the strength of the vector meson exchange interactions compared with the $\pi$ exchange interaction. However, the contribution of the vector meson exchange interactions may not be suppressed by the factor $(\Lambda^2-m_E^2)$ \cite{Ecker:1989yg}. In order to further test our conclusions, we use other kind of form factor $\mathcal{F}(q^2) = \Lambda^2/(\Lambda^2-q^2)$ to discuss the bound state properties \cite{Chen:2017vai}, which is not suppress the contribution of the vector meson exchange interactions. Here, we would like to mention that the cutoff values around 1.0 GeV in the form factor $\mathcal{F}(q^2,m_E^2) = (\Lambda^2-m_E^2)/(\Lambda^2-q^2)$ and 0.5 GeV in the form factor $\mathcal{F}(q^2) = \Lambda^2/(\Lambda^2-q^2)$ are the reasonable input parameters to study the hadronic molecular candidates \cite{Chen:2017vai}.
In Table~\ref{FFjg}, we give the bound state properties for the single $S$-wave $D_{1}D_{1}$ states with the form factors $\mathcal{F}(q^2,m_E^2) = (\Lambda^2-m_E^2)/(\Lambda^2-q^2)$ and $\mathcal{F}(q^2) = \Lambda^2/(\Lambda^2-q^2)$. For the $S$-wave $D_{1}D_{1}$ state with $I(J^P)=0(1^+)$, we can obtain the bound state solution by setting the cutoff values around 1.24 GeV in the form factor $\mathcal{F}(q^2,m_E^2)$ and 0.63 GeV in the form factor $\mathcal{F}(q^2)$, which reflects that the $S$-wave $D_{1}D_{1}$ state with $I(J^P)=0(1^+)$ can be viewed as a good doubly charmed molecular tetraquark candidate when using the two types of form factors. For the $S$-wave $D_{1}D_{1}$ states with $I(J^P)=1(0^+,\,2^+)$, we cannot find the bound state solutions with the two types of form factors until we increase the cutoff value to be around 2.5 GeV, which indicates that the $S$-wave $D_{1}D_{1}$ states with $I(J^P)=1(0^+,\,2^+)$ as the hadronic molecular candidates are no priority if we use the two types of form factors. Based on the analysis mentioned above, it is clear that the qualitative conclusion of the existence priority of the $S$-wave $D_{1}D_{1}$ hadronic molecular states does not change when increasing the contribution of the vector meson exchange interactions.
\renewcommand\tabcolsep{0.03cm}
\renewcommand{\arraystretch}{1.50}
\begin{table}[!htbp]
\caption{Bound state properties for the single $S$-wave $D_{1}D_{1}$ states with different form factors $\mathcal{F}(q^2,m_E^2) = (\Lambda^2-m_E^2)/(\Lambda^2-q^2)$ and $\mathcal{F}(q^2) = \Lambda^2/(\Lambda^2-q^2)$. Here, ``$\times$'' indicates no bound state solutions until we increase the cutoff parameter to be around 2.5 GeV.}\label{FFjg}
\centering
\begin{tabular}{c|ccc|ccc}\toprule[1.0pt]\midrule[1.0pt]
Form factors&\multicolumn{3}{c|}{$\mathcal{F}(q^2,m_E^2)$}&\multicolumn{3}{c}{$\mathcal{F}(q^2)$}\\\midrule[1.0pt]
$(I,J^P)$&$\Lambda(\rm{GeV})$ &$E(\rm {MeV})$ &$r_{\rm RMS}(\rm {fm})$&$\Lambda(\rm{GeV})$ &$E(\rm {MeV})$ &$r_{\rm RMS}(\rm {fm})$ \\\midrule[1.0pt]
$(1,0^+)$&$\times$&$\times$&$\times$    &$\times$&$\times$&$\times$\\
$(0,1^+)$&1.24&$-0.42$ &4.18        &0.63&$-0.35$ &4.60\\
         &1.35&$-6.80$&1.27         &0.82&$-5.62$ &1.49\\
         &1.45&$-19.50$ &0.81       &1.00&$-19.10$ &0.88\\
$(1,2^+)$&$\times$&$\times$&$\times$    &$\times$&$\times$&$\times$\\
\bottomrule[1.0pt]\midrule[1.0pt]
\end{tabular}
\end{table}}

{In the light pseudoscalar meson matrix $\mathbb{P}$,  there exists the $\eta-\eta^{\prime}$ mixing. In the following, we test the effect of the $\eta-\eta^{\prime}$ mixing in forming the bound states. If we consider the $\eta-\eta^{\prime}$ mixing, the light pseudoscalar meson matrix $\mathbb{P}$ in the effective Lagrangians is defined as
\begin{eqnarray}
\mathbb{P}=
\left(\begin{matrix} \frac{\pi^{0}}{\sqrt{2}}+\alpha\eta +\beta
\eta^{\prime }& \pi ^{-}\\
\pi ^{-} & -\frac{\pi ^{0}}{\sqrt{2}}+\alpha \eta +\beta \eta
^{\prime }
\end{matrix}\right),
\end{eqnarray}
where the parameters $\alpha$ and $\beta$ can be related to the mixing angle $\theta$ by
\begin{eqnarray}
\alpha=\frac{\cos\theta-\sqrt{2}\sin\theta}{\sqrt{6}},~~~~~~~\beta=\frac{\sin \theta+\sqrt{2}\cos \theta}{\sqrt{6}}.
\end{eqnarray}
Here, we use the mixing angle $\theta =-19.1^{\circ }$ \cite{Chen:2012nva}.
In Table~\ref{jgeep}, we present the numerical results both without and with considering the $\eta-\eta^{\prime}$ mixing for the single $S$-wave $D_{1}D_{1}$ state with $I(J^P)=0(1^+)$.  As presented in Table~\ref{jgeep}, if we take the same cutoff value without and with considering the $\eta-\eta^{\prime}$ mixing, we notice that the binding energies increase less than 4.0 MeV for the single $S$-wave $D_{1}D_{1}$ state with $I(J^P)=0(1^+)$. Thus, the $\eta-\eta^{\prime}$ mixing does not obviously affect the bound state properties, which is due to the contribution from the $\eta(\eta^{\prime})$ exchange interaction is small compared with the $\pi$ exchange interaction \cite{Chen:2016ryt}.
\renewcommand\tabcolsep{0.23cm}
\renewcommand{\arraystretch}{1.50}
\begin{table}[!htbp]
\caption{Bound state properties for the single $S$-wave $D_{1}D_{1}$ state with $I(J^P)=0(1^+)$ without and with considering the $\eta-\eta^{\prime}$ mixing.}\label{jgeep}
\centering
\begin{tabular}{c|cc|cc}\toprule[1.0pt]\toprule[1.0pt]
Cases&\multicolumn{2}{c|}{Without the $\eta-\eta^{\prime}$ mixing}&\multicolumn{2}{c}{With the $\eta-\eta^{\prime}$ mixing}\\\midrule[1.0pt]
$\Lambda(\rm{GeV})$ &$E(\rm {MeV})$ &$r_{\rm RMS}(\rm {fm})$ &$E(\rm {MeV})$ &$r_{\rm RMS}(\rm {fm})$  \\\midrule[1.0pt]
1.24&$-0.42$ &4.18&$-0.16$ &5.27\\
1.35&$-6.80$&1.27&$-5.21$&1.42\\
1.45&$-19.50$ &0.81&$-15.88$ &0.88\\
\bottomrule[1pt]\bottomrule[1pt]
\end{tabular}
\end{table}}

\subsection{The $S$-wave $D_1{D}_2^{\ast}$ system}\label{subsec32}

\renewcommand\tabcolsep{0.31cm}
\renewcommand{\arraystretch}{1.50}
\begin{table}[!htbp]
\caption{Bound state properties for the $S$-wave $D_1{D}_2^{\ast}$ system. The cutoff $\Lambda$, the binding energy $E$, and the root-mean-square radius $r_{RMS}$ are in units of $ \rm{GeV}$, $\rm {MeV}$, and $\rm {fm}$, respectively. Here, we label the major probability for the corresponding channels in a bold manner.}\label{D1D2}
\begin{tabular}{c|cccc}\toprule[1.0pt]\toprule[1.0pt]
\multicolumn{5}{c}{Single channel analysis}\\\midrule[1.0pt]
$I(J^{P})$&$\Lambda$ &$E$&$r_{\rm RMS}$ \\
\multirow{3}{*}{$0(1^{+})$} &1.04&$-0.29$&4.59   \\
                            &1.12&$-6.01$&1.33 \\
                            &1.20&$-19.53$&0.81 \\
\multirow{3}{*}{$0(2^{+})$} &1.45&$-0.42$&4.20   \\
                            &1.62&$-6.91$&1.27 \\
                            &1.78&$-20.03$&0.81 \\
\multirow{3}{*}{$1(3^{+})$} &2.26&$-0.50$&3.42   \\
                            &2.28&$-4.92$&1.08 \\
                            &2.30&$-11.50$&0.71 \\\midrule[1.0pt]
\multicolumn{5}{c}{$S$-$D$ wave mixing analysis}\\\midrule[1.0pt]
$I(J^{P})$&$\Lambda$&$E$&$r_{\rm RMS}$&$P({}^3\mathbb{S}_{1}/{}^3\mathbb{D}_{1}/{}^5\mathbb{D}_{1}/{}^7\mathbb{D}_{1})$  \\
\multirow{3}{*}{$0(1^{+})$} &1.04&$-0.42$&4.14&\textbf{99.89}/0.01/$o(0)$/0.10    \\
                            &1.12&$-6.37$&1.30&\textbf{99.83}/0.02/$o(0)$/0.15   \\
                            &1.20&$-20.05$&0.81&\textbf{99.85}/0.02/$o(0)$/0.13\\
$I(J^{P})$&$\Lambda$&$E$&$r_{\rm RMS}$&$P({}^5\mathbb{S}_{2}/{}^3\mathbb{D}_{2}/{}^5\mathbb{D}_{2}/{}^7\mathbb{D}_{2})$  \\
\multirow{3}{*}{$0(2^{+})$} &1.42&$-0.26$&4.83&\textbf{99.82}/$o(0)$/0.18/$o(0)$    \\
                            &1.59&$-6.10$&1.35&\textbf{99.62}/$o(0)$/0.38/$o(0)$   \\
                            &1.76&$-19.47$&0.82&\textbf{99.60}/$o(0)$/0.40/$o(0)$    \\
$I(J^{P})$&$\Lambda$&$E$&$r_{\rm RMS}$&$P({}^7\mathbb{S}_{3}/{}^3\mathbb{D}_{3}/{}^5\mathbb{D}_{3}/{}^7\mathbb{D}_{3})$  \\
\multirow{3}{*}{$0(3^{+})$} &2.47&$-0.26$&5.01&\textbf{98.81}/0.05/$o(0)$/1.14    \\
                            &2.49&$-0.30$&4.84&\textbf{98.75}/0.05/$o(0)$/1.20   \\
                            &2.50&$-0.33$&4.75&\textbf{98.72}/0.05/$o(0)$/1.23    \\
\multirow{3}{*}{$1(3^{+})$} &2.26&$-0.60$&3.17&\textbf{99.96}/0.01/$o(0)$/0.03    \\
                            &2.28&$-5.12$&1.06&\textbf{99.97}/0.01/$o(0)$/0.02   \\
                            &2.30&$-11.75$&0.70&\textbf{99.98}/0.01/$o(0)$/0.01   \\\midrule[1.0pt]
\multicolumn{5}{c}{Coupled channel analysis}\\\midrule[1.0pt]
$I(J^{P})$&$\Lambda$&$E$&$r_{\rm RMS}$&$P(D_{1}D_{2}^{*}/D_{2}^{*}D_{2}^{*})$  \\
\multirow{3}{*}{$0(1^{+})$}  &1.02&$-0.40$&4.18&\textbf{99.00}/1.00\\
                             &1.09&$-6.67$&1.24&\textbf{93.87}/6.13\\
                             &1.15&$-19.07$&0.79&\textbf{86.86}/13.14\\
\multirow{3}{*}{$0(3^{+})$}  &1.75&$-0.22$&4.98&\textbf{93.96}/6.14\\
                             &1.85&$-5.20$&1.25&\textbf{66.77}/33.23\\
                             &1.95&$-16.30$&0.73&48.78/51.22\\
\bottomrule[1pt]\bottomrule[1pt]
\end{tabular}
\end{table}

In Table~\ref{D1D2}, we present the obtained bound state properties for the $S$-wave $D_1{D}_2^{\ast}$ system by performing the single channel, $S$-$D$ wave mixing, and coupled channel analysis. For simplicity, we collect typical loosely bound state solutions for three groups, i.e.,
\begin{itemize}
  \item For the $D_1{D}_2^{\ast}$ system with $I(J^P)=0(1^+)$, the binding energy appears at the cutoff parameter around 1.04 GeV, when we take into account the single $|{}^3\mathbb{S}_1\rangle$ channel. When further adding the contribution from the $D$-wave channels, it is easy to form the $D_1{D}_2^{\ast}$ bound state with $I(J^P)=0(1^+)$. The probability of the $|{}^3\mathbb{S}_1\rangle$ channel is over 99\%, where the $S$-$D$ wave mixing effect can be ignored in forming the $S$-wave $D_1{D}_2^{\ast}$ bound state with $I(J^P)=0(1^+)$. By performing the coupled channel analysis with the $D_1{D}_2^{\ast}|{}^3\mathbb{S}_1\rangle$ and $D_{2}^{\ast}D_{2}^{\ast}|{}^3\mathbb{S}_1\rangle$ channels, we can obtain the loosely bound state solutions with the cutoff value $\Lambda=1.02$ GeV, and the probability for the $D_1{D}_2^{\ast}|{}^3\mathbb{S}_1\rangle$ channel is over 85\%.

  \item For the $D_1{D}_2^{\ast}$ system with $I(J^P)=0(2^+)$, the bound state solutions appear at the cutoff parameter larger than 1.45 GeV for the single channel case. In the $S$-$D$ wave mixing case, the corresponding cutoff parameter becomes smaller if getting the same binding energy, where the $D$-wave contribution is not obvious.

  \item For the $D_1{D}_2^{\ast}$ system with $I(J^P)=0(3^+)$, we can find a shallow binding energy with the cutoff value $\Lambda$ around 2.50 GeV after considering the $S$-$D$ wave mixing effect. However, taking this cutoff range, there is no bound state solutions for the single channel case. If we further consider the coupled channels like $D_1{D}_2^{\ast}|{}^7\mathbb{S}_3\rangle$ and $D_{2}^{\ast}D_{2}^{\ast}|{}^7\mathbb{S}_3\rangle$, the bound state solutions appear with the cutoff parameter around 1.75 GeV, and the $D_1{D}_2^{\ast}$ channel is the dominant component when the binding energy is less than 12.00 MeV. Thus, the coupled channel effect plays an important role for forming the $S$-wave $D_1{D}_2^{\ast}$ bound state with $I(J^P)=0(3^+)$.

  \item In the cutoff range $0.8<\Lambda<2.5$ GeV, we cannot find the bound state solutions for the $D_1{D}_2^{\ast}$ systems with $I(J^P)=1(1^+,\,2^+)$ even if we introduce the $S$-$D$ wave mixing effect and the coupled channel effect.

  \item For the $D_1{D}_2^{\ast}$ system with $I(J^P)=1(3^+)$, we obtain the loosely bound state solutions with the binding energy around several to several tens MeV, where the RMS radius is around several fm in the cutoff value larger than 2.20 GeV. Here, the bound state solutions are almost same for the single channel case and the $S$-$D$ wave mixing case.
\end{itemize}

To summarize, we can predict several possible doubly charmed molecular tetraquark candidates, such as the $S$-wave $D_1{D}_2^{\ast}$ states with $I(J^P)=0(1^+)$, $0(2^+)$, and $0(3^+)$. The $D_1{D}_2^{\ast}$ state with $I(J^P)=1(3^+)$ cannot be excluded as the possible doubly charmed molecular tetraquark candidate. In addition, we find the coupled channel effect plays an essential role in the formation of the $D_1{D}_2^{\ast}$ bound state with $I(J^P)=0(3^+)$.

\subsection{The $S$-wave $D_{2}^{\ast}D_{2}^{\ast}$ system}\label{subsec33}

\begin{figure}[!htbp]
\centering
\includegraphics[width=0.48\textwidth]{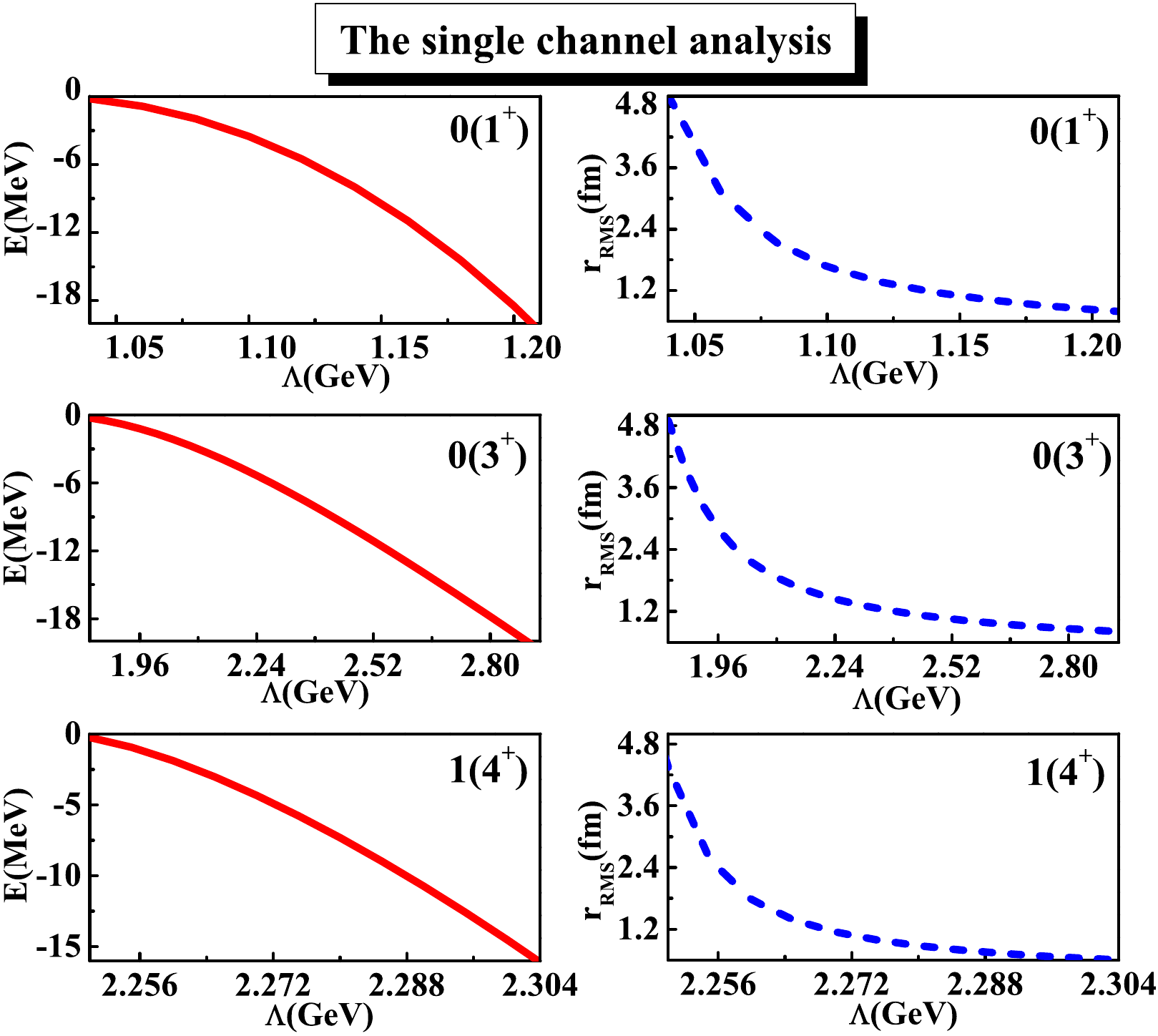}\\
\includegraphics[width=0.48\textwidth]{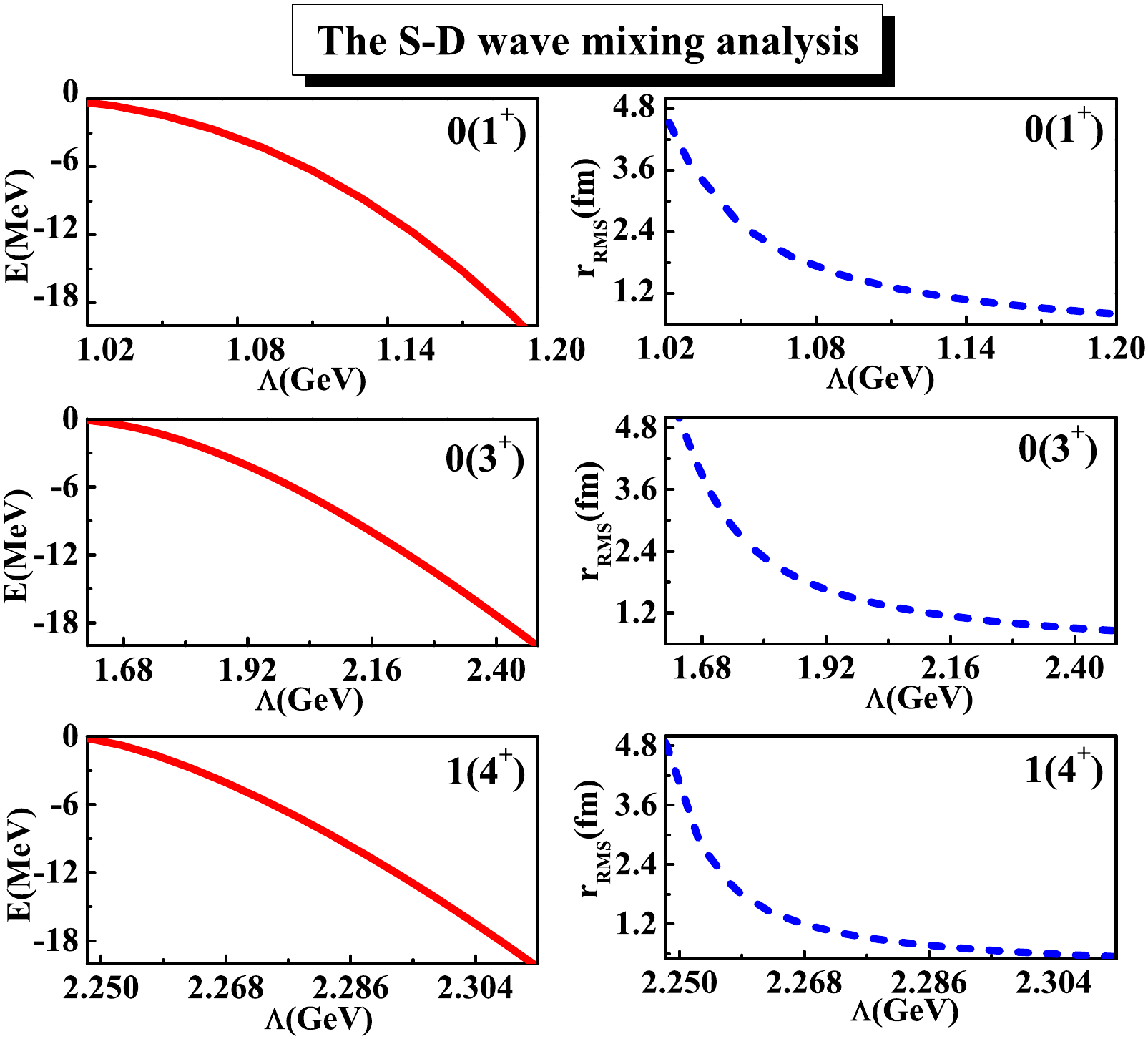}
\caption{The cutoff parameter $\Lambda$ dependence of the bound state solutions for the $S$-wave $D_{2}^{\ast}D_{2}^{\ast}$ system by performing the single channel and $S$-$D$ wave mixing analysis.}\label{D2D2}
\end{figure}

In Fig.~\ref{D2D2}, we present the relevant bound state properties for the $S$-wave $D_{2}^{\ast}D_{2}^{\ast}$ system by performing the single channel and $S$-$D$ wave mixing analysis.

For the $S$-wave $D_{2}^{\ast}D_{2}^{\ast}$ state with $I(J^P)=0(1^+)$,  there exists the bound state solutions with the small binding energy and the suitable RMS radius at the cutoff parameter around 1.04 GeV for the single channel analysis. Compared to the results for the single channel case, the bound state properties in the $S$-$D$ wave mixing analysis change slightly.

For the $S$-wave $D_{2}^{\ast}D_{2}^{\ast}$ state with $I(J^P)=0(3^+)$, the OBE effective interactions are a little weaker than those in the $S$-wave $D_{2}^{\ast}D_{2}^{\ast}$ state with $I(J^P)=0(1^+)$. When the cutoff value is taken to be larger than 1.87 GeV, we can obtain the bound state solutions for the single channel analysis. After considering the mixing effect of the $|{}^7\mathbb{S}_3\rangle$, $|{}^3\mathbb{D}_3\rangle$, and $|{}^7\mathbb{D}_3\rangle$ channels, there exists the loosely bound state solutions at the cutoff value $\Lambda>1.64 ~{\rm GeV}$, and the probabilities for the $D$-wave components are less than 2\%.

For the $S$-wave $D_{2}^{\ast}D_{2}^{\ast}$ states with $I(J^P)=1(0^+,\,2^+)$, their OBE effective potentials are weak attractive or repulsive by varying the cutoff parameter in the range of 0.80 to 2.50 GeV, these interactions are not strong enough to form the bound states. For the $D_{2}^{\ast}D_{2}^{\ast}$ state with $I(J^P)=1(4^+)$, we can obtain the loosely bound state solutions with the cutoff value around 2.25 GeV or even larger for both the single channel case and the $S$-$D$ wave mixing effect case, and the $S$-$D$ wave mixing effect plays a rather minor role to form the $S$-wave $D_{2}^{\ast}D_{2}^{\ast}$ bound state with $I(J^P)=1(4^+)$. Here, we need to specify that the cutoff parameter is a little larger than the reasonable value around $1.00$ GeV \cite{Tornqvist:1993ng,Tornqvist:1993vu}, and we conclude that the ${D}_2^{\ast}{D}_2^{\ast}$ state with $I(J^P)=1(4^+)$ may be viewed as the possible doubly charmed molecular tetraquark candidate.

Based on the above numerical results, the ${D}_2^{\ast}{D}_2^{\ast}$ states with $I(J^P)=0(1^+)$ and $0(3^+)$ can be good doubly charmed molecular tetraquark candidates, and the ${D}_2^{\ast}{D}_2^{\ast}$ state with $I(J^P)=1(4^+)$ as the possible doubly charmed molecular tetraquark candidate can be also acceptable.

{In the following, we discuss how the binding energies of the $S$-wave $D_1D_1$, $D_1{D}_2^{\ast}$, and ${D}_2^{\ast}{D}_2^{\ast}$ states can be affected by the $S$-wave $DD$, $D{D}^{\ast}$, and ${D}^{\ast}{D}^{\ast}$ states.
For example, when we restudy the bound state properties for the $S$-wave $D_{1}D_{1}$ state with $I(J^P)=0(1^+)$, we consider the $S$-wave $DD^*$, $D^*D^*$, and $D_{1}D_{1}$ scattering processes. As shown in the Table~\ref{jgsc}, we present the binding energies $E$ for the single $S$-wave $D_{1}D_{1}$ state with $I(J^P)=0(1^+)$ in the second column, and the mass gaps $\Delta M=M_{\text{resonance}}-M_{D_1D_1}$ between the mass of the obtained resonance and the $D_1D_1$ threshold after considering the $S$-wave $DD^*/D^*D^*/D_{1}D_{1}$ scattering processes in the third column. From the Table~\ref{jgsc}, we can find the masses for the $S$-wave $D_{1}D_{1}$ bound state with $I(J^P)=0(1^+)$ are almost the same with the masses for the obtained resonance when we take the same cutoff values. It indicates that the lower $DD^*$ and $D^*D^*$ states barely affect the binding energies for the $S$-wave $D_{1}D_{1}$ state with $I(J^P)=0(1^+)$, which is due to the mass threshold of the $D_1D_1$ channel is far from that of the ${D}^{(\ast)}{D}^{(\ast)}$ channels around 1.0 GeV.
\renewcommand\tabcolsep{0.54cm}
\renewcommand{\arraystretch}{1.50}
\begin{table}[!htbp]
\caption{Bound state properties for the single $S$-wave $D_{1}D_{1}$ state with $I(J^P)=0(1^+)$ without and with considering the scattering states.}\label{jgsc}
\centering
\begin{tabular}{c|c|c}\toprule[1.0pt]\toprule[1.0pt]
$\Lambda(\rm{GeV})$ &$E(\rm {MeV})$  &$M_{\text{resonance}}-M_{D_1D_1}(\rm {MeV})$   \\\midrule[1.0pt]
1.26&$-1.01$ &$-1.08$ \\
1.35&$-6.80$&$-6.90$\\
1.42&$-14.96$ &$-15.10$ \\
\bottomrule[1pt]\bottomrule[1pt]
\end{tabular}
\end{table}}

\section{Summary}\label{sec4}

In the past decades, the study of the exotic hadronic state has become an influential and attractive research field for the hadron physics. Benefited from the accumulation of experimental data with the high precision, the LHCb Collaboration reported a new structure $T_{cc}^+$ existing in the $D^0D^0\pi^+$ invariant mass spectrum \cite{LHCb:2021vvq}, where the $T_{cc}^+$ state can be regarded as the $DD^{\ast}$ doubly charmed molecular state with $J^{P}=1^{+}$ \cite{Li:2021zbw,Chen:2021vhg,Ren:2021dsi,Xin:2021wcr,Chen:2021tnn,Albaladejo:2021vln,Dong:2021bvy,Baru:2021ldu,Du:2021zzh,Kamiya:2022thy,Padmanath:2022cvl,Agaev:2022ast,Ke:2021rxd,Zhao:2021cvg,Deng:2021gnb,Santowsky:2021bhy,
Dai:2021vgf,Feijoo:2021ppq}. This new observation provides us a good opportunity to construct the family of the doubly charmed molecular tetraquarks.

In this work, we systematically study the $S$-wave interactions between a pair of charmed mesons in the $T$-doublet, where we can predict possible doubly charmed molecular tetraquarks. For the interactions between charmed meson pair in the $T$-doublet, we consider both the long-range contribution from the pseudoscalar meson exchange and the short-range and medium-range contributions from the vector and scalar meson exchanges. In the realistic calculation, we explore the roles of the single channel, the $S$-$D$ wave mixing effect, and the coupled channel effect to form the possible doubly charmed molecular tetraquark states, simultaneously.

Our numerical results show that the $S$-wave $D_1D_1$ state with $I(J^P)=0(1^+)$, the $S$-wave $D_1{D}_2^{\ast}$ states with $I(J^P)=0(1^+,\,2^+,\,3^+)$, and the $S$-wave ${D}_2^{\ast}{D}_2^{\ast}$ states with $I(J^P)=0(1^+,\,3^+)$ can be recommended as the prime doubly charmed molecular tetraquark candidates, which are consistent with the theoretical predictions in Ref. \cite{Dong:2021bvy}. {From the numerical results, we also can find that the binding properties of these most promising doubly charmed molecular tetraquark candidates are not significantly dependent on the cutoff values.} Meanwhile, the $S$-wave $D_1D_1$ state with $I(J^P)=1(2^+)$, the $S$-wave $D_1{D}_2^{\ast}$ state with $I(J^P)=1(3^+)$, and the $S$-wave ${D}_2^{\ast}{D}_2^{\ast}$ state with $I(J^P)=1(4^+)$ may be the secondary doubly charmed molecular tetraquark candidates. Their allowed decay modes include (i) two charmed mesons, (ii) a doubly charmed baryon plus a light anti-baryon, (iii) two charmed mesons plus one light meson, and (iv) two charmed mesons plus one photon.

Experimental search for these predicted doubly charmed molecular tetraquark candidates is an interesting and important research topic. With the accumulation of the Run II and Run III data \cite{Bediaga:2018lhg}, the LHCb Collaboration has the potential to hunt for these predicted doubly charmed molecular tetraquarks.

\section*{ACKNOWLEDGMENTS}

F. L. Wang would like to thank J. Z. Wang for very helpful discussions. This work is supported by the China National Funds for Distinguished Young Scientists under Grant No. 11825503, National Key Research and Development Program of China under Contract No. 2020YFA0406400, the 111 Project under Grant No. B20063, and the National Natural Science Foundation of China under Grant No. 12047501.

\appendix
\section{The operator matrix elements $\langle f|\mathcal{O}_k^{(\prime)}|i\rangle_{[J]}$}\label{app01}
For the operators $\mathcal{O}_k^{(\prime)}\,(k=1,\cdot\cdot\cdot,11)$, they should be sandwiched by the relevant spin-orbit wave functions $|{}^{2S+1}L_{J}\rangle$. In Table~\ref{matrix}, we present the operator matrix elements $\langle f|\mathcal{O}_k^{(\prime)}|i\rangle_{[J]}$ \cite{Chen:2015add}.
\renewcommand\tabcolsep{0.18cm}
\renewcommand{\arraystretch}{1.50}
\begin{table*}[htbp]
  \caption{The operator matrix elements $\langle f|\mathcal{O}_k^{(\prime)}|i\rangle_{[J]}$ for the operators $\mathcal{O}_k^{(\prime)}\,(k=1,\cdot\cdot\cdot,11)$ in the effective potentials.}\label{matrix}
\begin{tabular}{c|lll}\toprule[1pt]\toprule[1pt]
\multicolumn{1}{c|}{$\langle f|\mathcal{O}_k^{(\prime)}|i\rangle$}&\multicolumn{3}{c}{$\langle f|\mathcal{O}_k^{(\prime)}|i\rangle_{[J]}$}\\\midrule[1.0pt]
$\langle D_{1}D_{1}|\mathcal{O}_{1}|D_{1}D_{1}\rangle$
&${\rm{Diag}(1,1)}_{[0]}$ & ${\rm{Diag}(1,1)}_{[1]}$ & ${\rm{Diag}(1,1,1)}_{[2]}$\\
$\langle D_{1}D_{1}|\mathcal{O}_{2}|D_{1}D_{1}\rangle$
&${\rm{Diag}(2,-1)}_{[0]}$ & ${\rm{Diag}(1,1)}_{[1]}$ & ${\rm{Diag}(-1,2,-1)}_{[2]}$\\
$\langle D_{1}D_{1}|\mathcal{O}_{3}|D_{1}D_{1}\rangle$
&{$\left(\begin{array}{cc}0  &\sqrt{2}\\
                             \sqrt{2}  &2\end{array}\right)_{[0]}$}
                            &{$\left(\begin{array}{cc}0           &-\sqrt{2}\\
                                     -\sqrt{2}   &1 \end{array}\right)_{[1]}$}
                            &{$\left(\begin{array}{ccc}
                                0    &\sqrt{\frac{2}{5}}    &-\sqrt{\frac{14}{5}}\\
                                \sqrt{\frac{2}{5}}    &0     &-\frac{2}{\sqrt{7}}\\
                                -\sqrt{\frac{14}{5}}    &-\frac{2}{\sqrt{7}}    &-\frac{3}{7}
                                \end{array}\right)_{[2]}$}\\
$\langle D_{1}D_{2}^{\ast}|\mathcal{O}_{4}|D_{1}D_{2}^{\ast}\rangle$
&${\rm{Diag}(1,1,1,1)}_{[1]}$ & ${\rm{Diag}(1,1,1,1)}_{[2]}$ & ${\rm{Diag}(1,1,1,1)}_{[3]}$\\
$\langle D_{2}^{\ast}D_{1}|\mathcal{O}_{4}^{\prime}|D_{2}^{\ast}D_{1}\rangle$
&${\rm{Diag}(1,1,1,1)}_{[1]}$ & ${\rm{Diag}(1,1,1,1)}_{[2]}$ & ${\rm{Diag}(1,1,1,1)}_{[3]}$\\
$\langle D_{1}D_{2}^{\ast}|\mathcal{O}_{5}|D_{1}D_{2}^{\ast}\rangle$
&${\rm{Diag}(\frac{3}{2},\frac{3}{2},\frac{1}{2},-1)}_{[1]}$ & ${\rm{Diag}(\frac{1}{2},\frac{3}{2},\frac{1}{2},-1)}_{[2]}$ & ${\rm{Diag}(-1,\frac{3}{2},\frac{1}{2},-1)}_{[3]}$\\
$\langle D_{2}^{\ast}D_{1}|\mathcal{O}_{5}^{\prime}|D_{2}^{\ast}D_{1}\rangle$
&${\rm{Diag}(\frac{3}{2},\frac{3}{2},\frac{1}{2},-1)}_{[1]}$ & ${\rm{Diag}(\frac{1}{2},\frac{3}{2},\frac{1}{2},-1)}_{[2]}$ & ${\rm{Diag}(-1,\frac{3}{2},\frac{1}{2},-1)}_{[3]}$\\
$\langle D_{1}D_{2}^{\ast}|\mathcal{O}_{6}|D_{1}D_{2}^{\ast}\rangle$
&{$\left(\begin{array}{cccc}
                    0 & \frac{3}{5 \sqrt{2}} & \sqrt{\frac{6}{5}} & \frac{\sqrt{\frac{21}{2}}}{5} \\
                    \frac{3}{5 \sqrt{2}} & -\frac{3}{10} & \sqrt{\frac{3}{5}} & -\frac{\sqrt{\frac{3}{7}}}{5} \\
                    \sqrt{\frac{6}{5}} & \sqrt{\frac{3}{5}} & \frac{1}{2} & \frac{2}{\sqrt{35}} \\
                    \frac{\sqrt{\frac{21}{2}}}{5} & -\frac{\sqrt{\frac{3}{7}}}{5} & \frac{2}{\sqrt{35}} & \frac{48}{35}
                   \end{array}\right)_{[1]}$}
                            &{$\left(\begin{array}{cccc}
                    0  & -\frac{3 \sqrt{2}}{5} & -\sqrt{\frac{7}{10}} & \frac{\sqrt{7}}{5} \\
                    -\frac{3 \sqrt{2}}{5}  & \frac{3}{10} & \frac{3}{\sqrt{35}} & -\frac{3 \sqrt{\frac{2}{7}}}{5} \\
                    -\sqrt{\frac{7}{10}}  & \frac{3}{\sqrt{35}} & -\frac{3}{14} & \frac{4 \sqrt{\frac{2}{5}}}{7} \\
                    \frac{\sqrt{7}}{5}  & -\frac{3 \sqrt{\frac{2}{7}}}{5} & \frac{4 \sqrt{\frac{2}{5}}}{7} & \frac{12}{35}
                   \end{array}\right)_{[2]}$}
                            &{$\left(\begin{array}{cccc}
                    0 & \frac{3}{5 \sqrt{2}} & -\frac{1}{\sqrt{5}} & -\frac{4 \sqrt{3}}{5} \\
                    \frac{3}{5 \sqrt{2}} & -\frac{3}{35} & -\frac{6 \sqrt{\frac{2}{5}}}{7} & -\frac{6 \sqrt{6}}{35} \\
                    -\frac{1}{\sqrt{5}} & -\frac{6 \sqrt{\frac{2}{5}}}{7} & -\frac{4}{7} & \frac{\sqrt{\frac{3}{5}}}{7} \\
                    -\frac{4 \sqrt{3}}{5} & -\frac{6 \sqrt{6}}{35} & \frac{\sqrt{\frac{3}{5}}}{7} & -\frac{22}{35}
                   \end{array}\right)_{[3]}$}\\
$\langle D_{2}^{\ast} D_{1}|\mathcal{O}_{6}^{\prime}|D_{2}^{\ast}D_{1}\rangle$
&{$\left(\begin{array}{cccc}
                    0 & \frac{3}{5 \sqrt{2}} & -\sqrt{\frac{6}{5}} & \frac{\sqrt{\frac{21}{2}}}{5} \\
                    \frac{3}{5 \sqrt{2}} & -\frac{3}{10} & -\sqrt{\frac{3}{5}} & -\frac{\sqrt{\frac{3}{7}}}{5} \\
                    -\sqrt{\frac{6}{5}} & -\sqrt{\frac{3}{5}} & \frac{1}{2} & -\frac{2}{\sqrt{35}} \\
                    \frac{\sqrt{\frac{21}{2}}}{5} & -\frac{\sqrt{\frac{3}{7}}}{5} & -\frac{2}{\sqrt{35}} & \frac{48}{35}
                   \end{array}\right)_{[1]}$}
                             &{$\left(\begin{array}{cccc}
                    0  & \frac{3 \sqrt{2}}{5} & -\sqrt{\frac{7}{10}} & -\frac{\sqrt{7}}{5} \\
                    \frac{3 \sqrt{2}}{5}  & \frac{3}{10} & -\frac{3}{\sqrt{35}} & -\frac{3 \sqrt{\frac{2}{7}}}{5} \\
                    -\sqrt{\frac{7}{10}} & -\frac{3}{\sqrt{35}} & -\frac{3}{14} & -\frac{4 \sqrt{\frac{2}{5}}}{7} \\
                    -\frac{\sqrt{7}}{5} & -\frac{3 \sqrt{\frac{2}{7}}}{5} & -\frac{4 \sqrt{\frac{2}{5}}}{7} & \frac{12}{35}
                   \end{array}\right)_{[2]}$}
                             &{$\left(\begin{array}{cccc}
                    0 & \frac{3}{5 \sqrt{2}} & \frac{1}{\sqrt{5}} & -\frac{4 \sqrt{3}}{5} \\
                    \frac{3}{5 \sqrt{2}} & -\frac{3}{35} & \frac{6 \sqrt{\frac{2}{5}}}{7} & -\frac{6 \sqrt{6}}{35} \\
                    \frac{1}{\sqrt{5}} & \frac{6 \sqrt{\frac{2}{5}}}{7} & -\frac{4}{7} & -\frac{\sqrt{\frac{3}{5}}}{7} \\
                    -\frac{4 \sqrt{3}}{5} & -\frac{6 \sqrt{6}}{35} & -\frac{\sqrt{\frac{3}{5}}}{7} & -\frac{22}{35}
                   \end{array}\right)_{[3]}$}\\
$\langle D_{2}^{\ast}D_{1}|\mathcal{O}_{7}|D_{1}D_{2}^{\ast}\rangle$
&${\rm{Diag}(\frac{1}{6},\frac{1}{6},\frac{1}{2},1)}_{[1]}$ & ${\rm{Diag}(\frac{1}{2},\frac{1}{6},\frac{1}{2},1)}_{[2]}$ & ${\rm{Diag}(1,\frac{1}{6},\frac{1}{2},1)}_{[3]}$\\
$\langle D_{1}D_{2}^{\ast}|\mathcal{O}_{7}^{\prime}|D_{2}^{\ast} D_{1}\rangle$
&${\rm{Diag}(\frac{1}{6},\frac{1}{6},\frac{1}{2},1)}_{[1]}$ & ${\rm{Diag}(\frac{1}{2},\frac{1}{6},\frac{1}{2},1)}_{[2]}$ & ${\rm{Diag}(1,\frac{1}{6},\frac{1}{2},1)}_{[3]}$\\
$\langle D_{2}^{\ast}D_{1}|\mathcal{O}_{8}|D_{1}D_{2}^{\ast}\rangle$
&{$\left(\begin{array}{cccc}
                    0 & -\frac{23}{15 \sqrt{2}} & -2 \sqrt{\frac{2}{15}} & -\frac{\sqrt{\frac{7}{6}}}{5} \\
                    -\frac{23}{15 \sqrt{2}} & \frac{23}{30} & -\frac{2}{\sqrt{15}} & \frac{1}{5 \sqrt{21}} \\
                    2 \sqrt{\frac{2}{15}} & \frac{2}{\sqrt{15}} & \frac{1}{2} & -\frac{2}{\sqrt{35}} \\
                    -\frac{\sqrt{\frac{7}{6}}}{5} & \frac{1}{5 \sqrt{21}} & \frac{2}{\sqrt{35}} & \frac{24}{35}
                   \end{array}\right)_{[1]}$}
                             &{$\left(\begin{array}{cccc}
                    0  & -\frac{2 \sqrt{2}}{5} & -\sqrt{\frac{7}{10}} & -\frac{\sqrt{7}}{5} \\
                    \frac{2 \sqrt{2}}{5}  & -\frac{23}{30} & -\frac{2}{\sqrt{35}} & \frac{\sqrt{\frac{2}{7}}}{5} \\
                    -\sqrt{\frac{7}{10}}  & \frac{2}{\sqrt{35}} & -\frac{3}{14} & -\frac{4 \sqrt{\frac{2}{5}}}{7} \\
                    \frac{\sqrt{7}}{5}  & \frac{\sqrt{\frac{2}{7}}}{5} & \frac{4 \sqrt{\frac{2}{5}}}{7} & \frac{6}{35}
                   \end{array}\right)_{[2]}$}
                             &{$\left(\begin{array}{cccc}
                    0 & -\frac{1}{5 \sqrt{2}} & -\frac{1}{\sqrt{5}} & -\frac{2 \sqrt{3}}{5} \\
                    -\frac{1}{5 \sqrt{2}} & \frac{23}{105} & \frac{4 \sqrt{\frac{2}{5}}}{7} & \frac{2 \sqrt{6}}{35} \\
                    \frac{1}{\sqrt{5}} & -\frac{4 \sqrt{\frac{2}{5}}}{7} & -\frac{4}{7} & -\frac{\sqrt{\frac{3}{5}}}{7} \\
                    -\frac{2 \sqrt{3}}{5} & \frac{2 \sqrt{6}}{35} & \frac{\sqrt{\frac{3}{5}}}{7} & -\frac{11}{35}
                   \end{array}\right)_{[3]}$}\\
$\langle D_{1}D_{2}^{\ast}|\mathcal{O}_{8}^{\prime}|D_{2}^{\ast} D_{1}\rangle$
&{$\left(\begin{array}{cccc}
                    0 & -\frac{23}{15 \sqrt{2}} & 2 \sqrt{\frac{2}{15}} & -\frac{\sqrt{\frac{7}{6}}}{5} \\
                    -\frac{23}{15 \sqrt{2}} & \frac{23}{30} & \frac{2}{\sqrt{15}} & \frac{1}{5 \sqrt{21}} \\
                    -2 \sqrt{\frac{2}{15}} & -\frac{2}{\sqrt{15}} & \frac{1}{2} & \frac{2}{\sqrt{35}} \\
                    -\frac{\sqrt{\frac{7}{6}}}{5} & \frac{1}{5 \sqrt{21}} & -\frac{2}{\sqrt{35}} & \frac{24}{35}
                   \end{array}\right)_{[1]}$}
                            &{$\left(\begin{array}{cccc}
                    0 & \frac{2 \sqrt{2}}{5} & -\sqrt{\frac{7}{10}} & \frac{\sqrt{7}}{5} \\
                    -\frac{2 \sqrt{2}}{5} & -\frac{23}{30} & \frac{2}{\sqrt{35}} & \frac{\sqrt{\frac{2}{7}}}{5} \\
                    -\sqrt{\frac{7}{10}} & -\frac{2}{\sqrt{35}} & -\frac{3}{14} & \frac{4 \sqrt{\frac{2}{5}}}{7} \\
                    -\frac{\sqrt{7}}{5} & \frac{\sqrt{\frac{2}{7}}}{5} & -\frac{4 \sqrt{\frac{2}{5}}}{7} & \frac{6}{35}
                   \end{array}\right)_{[2]}$}
                            &{$\left(\begin{array}{cccc}
                    0 & -\frac{1}{5 \sqrt{2}} & \frac{1}{\sqrt{5}} & -\frac{2 \sqrt{3}}{5} \\
                    -\frac{1}{5 \sqrt{2}} & \frac{23}{105} & -\frac{4 \sqrt{\frac{2}{5}}}{7} & \frac{2 \sqrt{6}}{35} \\
                    -\frac{1}{\sqrt{5}} & \frac{4 \sqrt{\frac{2}{5}}}{7} & -\frac{4}{7} & \frac{\sqrt{\frac{3}{5}}}{7} \\
                    -\frac{2 \sqrt{3}}{5} & \frac{2 \sqrt{6}}{35} & -\frac{\sqrt{\frac{3}{5}}}{7} & -\frac{11}{35}
                   \end{array}\right)_{[3]}$}\\
$\langle D_{2}^{\ast}D_{2}^{\ast}|\mathcal{O}_{9}|D_{2}^{\ast}D_{2}^{\ast}\rangle$
&${\rm{Diag}(1,1)}_{[0]}$ & ${\rm{Diag}(1,1,1)}_{[1]}$ & ${\rm{Diag}(1,1,1,1)}_{[2]}$ \\
& ${\rm{Diag}(1,1,1)}_{[3]}$ & ${\rm{Diag}(1,1,1)}_{[4]}$\\
$\langle D_{2}^{\ast}D_{2}^{\ast}|\mathcal{O}_{10}|D_{2}^{\ast}D_{2}^{\ast}\rangle$
&${\rm{Diag}(\frac{3}{2},\frac{3}{4})}_{[0]}$ & ${\rm{Diag}(\frac{5}{4},\frac{5}{4},0)}_{[1]}$ & ${\rm{Diag}(\frac{3}{4},\frac{3}{4},\frac{3}{4},-1)}_{[2]}$ \\
& ${\rm{Diag}(0,\frac{5}{4},0)}_{[3]}$ & ${\rm{Diag}(-1,\frac{3}{4},-1)}_{[4]}$\\
$\langle D_{2}^{\ast}D_{2}^{\ast}|\mathcal{O}_{11}|D_{2}^{\ast}D_{2}^{\ast}\rangle$
&{$\left(\begin{array}{cc}
                                 0    &\frac{3\sqrt{\frac{7}{10}}}{2}    \\
                                 \frac{3\sqrt{\frac{7}{10}}}{2}      &\frac{15}{14}
                                 \end{array}\right)_{[0]}$}
                             &{$\left(\begin{array}{ccc}
                                  0 & -\frac{13}{10 \sqrt{2}} & \frac{3 \sqrt{7}}{10} \\
                                  -\frac{13}{10 \sqrt{2}} & \frac{13}{20} & -\frac{3}{5 \sqrt{14}} \\
                                  \frac{3 \sqrt{7}}{10} & -\frac{3}{5 \sqrt{14}} & \frac{36}{35}
                                 \end{array}\right)_{[1]}$}
                             &{$\left(\begin{array}{cccc}
                                  0 & \frac{3 \sqrt{\frac{7}{2}}}{10} & -\frac{3 \sqrt{\frac{5}{14}}}{2} & \frac{9}{5 \sqrt{14}} \\
                                  \frac{3 \sqrt{\frac{7}{2}}}{10} & 0 & -\frac{3}{2 \sqrt{5}} & 0 \\
                                  -\frac{3 \sqrt{\frac{5}{14}}}{2} & -\frac{3}{2 \sqrt{5}} & -\frac{45}{196} & -\frac{18}{49 \sqrt{5}} \\
                                  \frac{9}{5 \sqrt{14}} & 0 & -\frac{18}{49 \sqrt{5}} & 0
                                 \end{array}\right)_{[2]}$}\\
{}&{$\left(\begin{array}{ccc}
                                0    &\frac{3\sqrt{3}}{10}     &-\frac{3\sqrt{3}}{5}\\
                                \frac{3\sqrt{3}}{10}    &\frac{13}{70}      &-\frac{18}{35}\\
                                -\frac{3\sqrt{3}}{5}    &-\frac{18}{35}      &-\frac{33}{70}
                               \end{array}\right)_{[3]}$}
                         &{$\left(\begin{array}{ccc}
                                0    &\frac{3}{\sqrt{70}}       &-\sqrt{\frac{11}{7}}\\
                               \frac{3}{\sqrt{70}}  &\frac{15}{49}  &-\frac{3\sqrt{\frac{55}{2}}}{49}\\
                                -\sqrt{\frac{11}{7}} &-\frac{3\sqrt{\frac{55}{2}}}{49}  &\frac{65}{98}
                             \end{array}\right)_{[4]}$}\\
\bottomrule[1pt]\bottomrule[1pt]
\end{tabular}
\end{table*}

\end{document}